\documentclass[]{aa}  
\usepackage[nottoc]{tocbibind}
\usepackage{enumerate}
\usepackage{float}
\usepackage{titlesec}
\usepackage{afterpage}
\usepackage{amssymb,amsmath}
\usepackage[]{graphicx} \graphicspath{ {Fig/} }
\usepackage{caption}
\usepackage{subfig}
\usepackage[]{xcolor}
\usepackage{array}
\usepackage{booktabs, multirow}
\usepackage[colorlinks,linkcolor=black,urlcolor=blue,citecolor=blue]{hyperref}
\makeatletter
\renewcommand*\aa@pageof{, page \thepage{} of \pageref*{LastPage}}
\makeatother\usepackage{natbib}

\usepackage{url}
\usepackage[utf8]{inputenc}
\usepackage{titlesec}
\usepackage{amssymb,amsmath}
\usepackage{verbatim}
\usepackage{graphicx}
\usepackage{enumitem}
\usepackage{color}
\usepackage{sidecap}
\usepackage{graphicx, caption}
\usepackage{caption}
\usepackage{cuted}
\renewcommand{\theenumi}{(\roman{enumi})}%

\usepackage{txfonts}
\begin{document}

   \title{The Back-in-time Void Finder: dynamical identification of cosmic voids through optimal transport reconstruction}

   \author{S. Sartori\inst{1,2}, S. Contarini\inst{2,3}, E. Sarpa\inst{4,5,6}, G. Degni\inst{1}, F. Marulli\inst{2,7,8}, S. Escoffier\inst{1}, L. Moscardini\inst{2,7,8}}

   \institute{Aix Marseille Université, CNRS/IN2P3, CPPM, Marseille, France \and 
   Dipartimento di Fisica e Astronomia “Augusto Righi” - Alma Mater Studiorum Università di Bologna, via Piero Gobetti 93/2, 40129 Bologna, Italy \and 
   Max Planck Institute for Extraterrestrial Physics, Giessenbachstrasse 1, 85748 Garching, Germany \and 
   SISSA, International School for Advanced Studies, Via Bonomea 265, 34136 Trieste, Italy \and 
   ICSC - Centro Nazionale di Ricerca in High Performance Computing, Big Data e Quantum Computing, Via Magnanelli 2, 40033 Casalecchio di Reno (BO), Italy \and 
   INAF - Osservatorio Astronomico di Trieste, Via G. B. Tiepolo 11, 34143 Trieste, Italy \and 
   INAF - Osservatorio di Astrofisica e Scienza dello Spazio di Bologna, via Piero Gobetti 93/3, 40129 Bologna, Italy \and 
   INFN - Sezione di Bologna, viale Berti Pichat 6/2, 40127 Bologna, Italy}


\abstract{Cosmic voids have increasingly emerged as a powerful cosmological probe. However, their large spatial extent and intrinsically underdense environments make their identification highly sensitive to shot noise, redshift-space distortions (RSD), and observational systematics, particularly for topological and density-based void definitions. 

We introduce the \texttt{Back-In-Time Void Finder} (\texttt{BitVF}), a novel dynamical and physically motivated algorithm that identifies cosmic voids as regions of negative divergence of the Lagrangian displacement field reconstructed from the present-day tracer distribution. The reconstruction relies on an optimized discrete optimal transport algorithm that recovers the backward-in-time dynamics of tracers, naturally accounting for tracer bias without relying on cosmological assumptions. 

We validate \texttt{BitVF} against the widely used topological void finder \texttt{REVOLVER} using high-resolution N-body simulations, showing that it produces void catalogs with smoother and more physically motivated density profiles, as well as abundances that are more stable under tracer subsampling and shot noise. We further apply it to realistic DESI-like mock light-cone galaxy catalogs, demonstrating that it intrinsically mitigates redshift-space systematic effects, preserving real-space void size functions more faithfully than topological methods. Modeling RSD, the reconstruction can be combined with a fiducial cosmology and an assumed tracer bias within a bias-corrected Kaiser framework, yielding reconstructed-space void catalogs consistent with real-space statistics across redshift. Its performance is characterized as a function of the main internal parameters, showing an optimal balance between accuracy, computational efficiency, and applicability to stage IV galaxy surveys. \texttt{BitVF} will be publicly released within the \texttt{CosmoBolognaLib}.}


   \authorrunning{S. Sartori et al.}
   \titlerunning{The Back-in-time void finder}
   \maketitle
%

\section{Introduction}

The large-scale structure of the Universe emerges as a complex network of filaments, walls, and clusters enclosing vast underdense regions known as cosmic voids. Although voids contain only a small fraction of the total matter content, they dominate the cosmic volume and play a key role in shaping the geometry and dynamics of the cosmic web \citep{Cautun2014,KugelvdW2024}.  

The presence of such large underdensities has been recognized since the first redshift surveys \citep{G&T1978, joeveer1978}, which revealed a filamentary pattern of galaxy clustering separated by apparently empty regions extending over tens of megaparsecs. Since then, voids have been extensively investigated in both simulations and observations, and are now well established as a powerful cosmological probe (see \citealt{pisani2019cosmic, Moresco2022,CaiNeyrinck2025} reviews and references therein).  

Nevertheless, a general consensus on the definition of cosmic voids has not yet been reached. Given their intrinsically complex nature, voids can be characterized according to \textit{topological}, \textit{density-based}, or \textit{dynamical} criteria \citep{L&W2010}, emphasizing different aspects of their structure and evolution, such as their underdense environment, typical spherical shapes, and the coherent mass outflows that govern their dynamics. 

While the absence of a universal definition is not problematic per se, as long as the adopted one is consistent with the theoretical modeling assumed when performing the analysis, topological and density-based approaches are particularly susceptible to shot noise. This limitation arises from the sparse sampling of the underlying mass 
distribution by tracers in underdense regions and is further amplified by the typically higher linear galaxy bias within voids compared to the rest of the 
Universe \citep{Pollina2017, Contarini2019}. Moreover, void identification in redshift space is strongly affected by distortions of void shapes induced by redshift-space distortions (RSD) and the Alcock–Paczynski (AP) effect \citep{AP1979}, 
which break the one-to-one mapping between real- and redshift-space voids, systematically shift the identified centers along the line-of-sight (LOS), and introduce significant selection effects \citep{Cai2016, Nadathur2019, Correa2022b}.

To overcome these limitations, we introduce a novel void finder based on a dynamical definition: the \texttt{Back-in-time Void Finder} (\texttt{BitVF}). In this framework, voids are identified as regions where the divergence of the displacement field, reconstructed back in time from the present-day Eulerian space to the primordial Lagrangian conditions, is negative ($\nabla_\mathbf{q} \cdot \boldsymbol{\Psi} < 0$). In a Lagrangian description, such regions correspond to basins of local mass outflow from an initially homogeneous matter distribution, providing a physically motivated criterion for void identification.

The displacement field $\boldsymbol{\Psi}$ can in principle be obtained from any reconstruction algorithm mapping present-day tracer distributions back to a uniform primordial configuration, as predicted by the cosmological principle. In this work, we employ a reconstruction method based on optimal transport (OT; see, e.g, \citealt{Brenier1991,BenamouBrenier2000,Frisch2002,Brenier2003,Nikakhtar2022,Nikakhtar2023,Nikakhtar2024} for introductory works and some cosmological applications of OT), building on the Particle Interchange Zel'dovich Approximation (\texttt{PIZA}) pairwise-swapping concept (\citealt{piza1997}, further refined by \citealt{Elyiv2015}). The reconstructed displacement field naturally encodes global dynamical information. Unlike direct density estimates, where each point in space is informed only by the local tracer density and is therefore highly sensitive to shot noise, every location in the displacement field $\boldsymbol{\Psi}$ reflects the coherent large-scale mass flow. This non-locality makes $\boldsymbol{\Psi}$ intrinsically more robust to sampling fluctuations than the density field. This property, together with the intrinsic smoothness of the field, mitigates shot noise and enables a void definition rooted in dynamics rather than in ad hoc prescriptions.

The OT method naturally incorporates tracer bias into the reconstructed field and does not rely on a fiducial cosmology, unless one intends to model RSD to identify voids in reconstructed space. Its validity extends as long as large-scale motions remain predominantly irrotational and single-stream, conditions well satisfied in underdense regions \citep{Brenier1991}. The resulting framework is optimized for both real and simulated data, and can explicitly account for survey geometry, angular masks, and completeness corrections through the use of a random catalog mimicking the survey selection functions \citep[see, e.g,][]{Eisenstein2007, White2015, Sarpa2021}. Moreover, when a fiducial cosmology and the tracer bias are specified, the method can also correct for RSD, effectively mapping reconstructed voids into real space, enhancing the interpretability of the resulting catalogs.

The paper is organized as follows. In Sect. \ref{sec:theory}, we present the theoretical framework of our approach. In Sect. \ref{sec:simulations} we describe the simulations used in this work. Section \ref{sec:bitvf} introduces our new dynamical void finder and details its implementation. Additionally, OT reconstructed velocities are tested in Appendix \ref{app:velcomp}, while parameter-sensitivity tests are provided in Appendix \ref{app:impact}. We show a first application to simulation boxes in Sect. \ref{sec:tests}, where we validate the identified voids by comparing them with a topological void sample identified in the same simulation, and we assess the robustness of our results against variations in tracer sparsity. In Sect. \ref{sec:lightcone}, we apply the method to light-cone mock catalogs, examining void properties in real space, redshift space, and reconstructed real space. Moreover, we introduce a OT-oriented modification of the \citet{Kaiser1987} formalism, expanded in Appendix \ref{app:otderiv}. Finally, in Sect. \ref{sec:conclusions} we summarize our findings and outline prospects for future applications.

The \texttt{BitVF} code, along with the OT reconstruction and all the cosmological tools used in this work, are made publicly available within the \texttt{CosmoBolognaLib}\footnote[1]{\url{https://gitlab.com/federicomarulli/CosmoBolognaLib}} package \citep{cbl}, a comprehensive set of \texttt{C++} and \texttt{Python} free software for cosmological computations.

\section{Theoretical background}
\label{sec:theory}

In the Lagrangian description of cosmic structure formation, the evolution of matter is modeled as a mapping between the initial positions $\mathbf{q}$ of mass elements, drawn from an initially homogeneous distribution, and their comoving coordinates $\mathbf{x}$ at a given cosmic time \citep{Peebles1989,Buchert1992,Bernardeau2002}. At linear order this mapping can be written as
\begin{equation}
   \mathbf{x}(\mathbf{q},t) = \mathbf{q} + \boldsymbol{\Psi}(\mathbf{q},t) \, ,
\end{equation}
where $\boldsymbol{\Psi}(\mathbf{q},t)$ is the Lagrangian displacement field. Within the Zel’dovich approximation \citep{Zeldovich1970}, $\boldsymbol{\Psi}$ is irrotational and derives from the gradient of a scalar potential, guaranteeing a unique mapping prior to shell crossing \citep{Brenier1991, Brenier2003}.

The evolution of the matter density field follows from the Jacobian of the transformation can be reduced in the linear regime as:
\begin{equation}
1 + \delta(\mathbf{x}, t)
= \left| \frac{\partial \mathbf{x}(t)}{\partial \mathbf{q}} \right|^{-1}
= \left| \frac{\partial \big[\mathbf{q} + \boldsymbol{\Psi}(\mathbf{q}, t)\big]}{\partial \mathbf{q}} \right|^{-1}
\simeq 1 - \nabla_{\mathbf{q}} \cdot \boldsymbol{\Psi}(\mathbf{q}, t) \, .
\end{equation}

This relation directly associates overdensities with negative divergence, and underdensities with positive divergence of the displacement field.  

In our approach, however, the displacement is reconstructed backward in time, namely
\begin{equation}
\boldsymbol{\Psi} = \mathbf{x} - \mathbf{q} \, ,
\end{equation}
where tracers are transported from their observed Eulerian positions back to an initially homogeneous Lagrangian distribution. Under this convention the sign of the divergence is inverted: voids correspond to regions with
\begin{equation}
\nabla_\mathbf{q} \cdot \boldsymbol{\Psi} < 0 \, ,
\end{equation}
corresponding to the expected mass outflow from underdense regions when evolved forward in time. This sign convention is therefore central to our dynamical void definition.  

Since the Zel’dovich approximation remains valid in underdense environments, where multistreaming is rare and the displacement field well described by a scalar potential, this criterion provides a robust and physically motivated basis for cosmic void identification \citep{SahniShandarin1996, L&W2010, McQuinnWhite2016}.

\subsection{Optimal transport approach to the Lagrangian displacement reconstruction}
\label{sec:ottheory}

The reconstruction of the displacement field \(\mathbf{\Psi}\) can be cast in the OT framework, which provides a mathematically rigorous way to 
determine the most efficient mapping between two mass distributions (for an extensive review on the topic, see \citealt{Villani2008}). The fundamental principle of OT relies on minimizing a global transport cost $\mathcal{C}$, or \textit{Monge cost function} \citep{Monge1781}, corresponding in our context to the total squared displacement required to map the present-day inhomogeneous tracer distribution back to an initially uniform mass configuration:
\begin{equation}
   \mathcal{C} = \int_{\mathbb{R}^3} \| T(\mathbf{q}) - \mathbf{q} \|^2 \, \mathrm{d}^3\mathbf{q} \, ,
\end{equation}
where \(T: \mathbf{q} \rightarrow \mathbf{x}\) denotes the bijective map between Lagrangian and Eulerian coordinates. In the discrete form, relevant for our case, this reduces to minimizing the sum of squared distances between matched particles under one-to-one assignment constraints \citep{Brenier2003}. This formulation yields a convex optimization problem with a unique solution under mild regularity assumptions \citep{Brenier1991}.  

In this sense, the OT reconstruction provides a practical realization of the Lagrangian mapping, with the additional benefit of being directly applicable to discrete tracer catalogs. 
Crucially, the OT-based reconstruction does not rely on any cosmological assumptions: it purely determines the geometric optimal mapping between the observed tracer positions and the initial uniform mass distribution. Cosmological parameters only enter when, after the reconstruction process, the reconstructed displacement field is interpreted physically, for instance when 
computing peculiar velocities via linear theory or correcting for the effect of RSD.

Instead of solving the full variational problem, which is computationally demanding, we adopt an approximated but efficient implementation inspired by the \texttt{PIZA} method \citep{piza1997}. In this scheme, the optimal mapping is obtained iteratively by swapping Eulerian-Lagrangian position pairs so as to reduce the global transport cost. This discrete algorithm is well-suited to the practical reconstruction of galaxy and dark matter catalogs, while still retaining the 
essential OT properties.  

An additional advantage of the OT approach is that it naturally incorporates tracer bias. 
The displacement field obtained by applying OT to a tracer population with linear bias 
\(b\), \(\boldsymbol{\Psi}_\mathrm{tr}\), is related to the displacement field that 
would be reconstructed from the underlying matter distribution, 
\(\boldsymbol{\Psi}_\mathrm{m}\), through the standard linear-bias model. 
In terms of their divergences one has
\begin{equation}
\label{deltabiased}
\nabla \cdot \boldsymbol{\Psi}_\mathrm{tr} \simeq - \delta_\mathrm{tr} 
= - b \, \delta_\mathrm{m} \simeq b \, \nabla \cdot \boldsymbol{\Psi}_\mathrm{m} \, ,
\end{equation}
which implies the simple scaling
\begin{equation}
\boldsymbol{\Psi}_\mathrm{m} \simeq \frac{1}{b} \, \boldsymbol{\Psi}_\mathrm{tr} \, .
\end{equation}
This guarantees that our displacement reconstruction is fully independent of any cosmological assumptions. Furthermore, as shown in Eq.~\eqref{deltabiased}, the density contrast of the biased tracer population can be directly derived from the divergence of the reconstructed displacement field.

\subsection{Reconstruction of real-space positions and redshift-space distortions}

When observing galaxies in redshift space, the tracer positions along the LOS are systematically shifted by the contribution of peculiar velocities to the observed redshift.
These shifts, the so-called RSD, introduce anisotropies in the observed large-scale structure and bias the inferred positions of underdense and overdense regions, afflicting our void samples and statistical measures \citep{Kaiser1987, Hamilton1998}.

The procedure to recover real-space positions from redshift-space observations, following \citet{Kaiser1987}, can be expressed in terms of the displacement 
field \(\mathbf{\Psi}\). For a continuous matter field, the mapping from redshift-space positions \(\mathbf{s}\) to real-space positions \(\mathbf{x}\) reads
\begin{equation}
    \label{eq:originalxs}
    \mathbf{x} = \mathbf{s} - f \, \mathbf{\Psi}_\parallel \, , \quad
    \mathbf{\Psi}_\parallel \equiv (\mathbf{\Psi} \cdot \hat{\mathbf{r}}) \hat{\mathbf{r}} \, ,
\end{equation}
where \(f \equiv \mathrm{d}\ln D / \mathrm{d}\ln a\) is the linear growth rate and the parallel component \(\mathbf{\Psi}_\parallel\) is defined along the LOS direction \(\hat{\mathbf{r}}\). Here, \(\mathbf{\Psi}\) is derived from the underlying matter density field and represents the true, unbiased displacement $\boldsymbol{\Psi}_\mathrm{m}$. This equation provides the standard linear-theory prescription to correct RSD in the absence of tracer bias.

In the Lagrangian framework, the peculiar velocity field is directly related to the displacement field through linear theory \citep{Peebles1980}:
\begin{equation}
\label{eq:velpsirel}
\mathbf{v}(\mathbf{x}) = a f H \,\boldsymbol{\Psi}(\mathbf{x}) \, ,
\end{equation}
where \(a\) is the scale factor, \(H\) the Hubble parameter, and \(f\) the linear growth rate of structure. 

In redshift space, the observed tracer positions $\mathbf{s}$ are related to the Lagrangian coordinates $\mathbf{q}$ via the total displacement $\boldsymbol{\Psi}^{(s)}$ \citep{Kaiser1987}:
\begin{equation}
\mathbf{s} = \mathbf{q} + \boldsymbol{\Psi}^{(s)}(\mathbf{q}) 
= \mathbf{q} + \boldsymbol{\Psi}(\mathbf{q}) + \frac{(\mathbf{v}(\mathbf{q})\cdot\hat{\mathbf{r}})\,\hat{\mathbf{r}}}{a H} 
= \mathbf{q} + \boldsymbol{\Psi}(\mathbf{q}) + f\,(\boldsymbol{\Psi}(\mathbf{q})\cdot\hat{\mathbf{r}})\,\hat{\mathbf{r}} \, ,
\end{equation}
where the last equality uses linear theory (Eq.~\ref{eq:velpsirel}) to express the LOS RSD contribution directly in terms of \(\boldsymbol{\Psi}\). Solving the OT problem in redshift space therefore yields \(\boldsymbol{\Psi}^{(s)}\), automatically incorporating both real-space evolution and RSD. The RSD component can be isolated to recover the real-space positions assuming a fiducial value of \(f\).

Moreover, recalling that OT maps the present-day Eulerian positions of the tracers onto a uniform distribution, when the displacement field is reconstructed from biased tracers, their linear bias \(b\) must be taken into account to recover the true tracer displacement, which is expected to be unbiased.

Consequently, both the peculiar velocities and the mapping from redshift to real space derived from the reconstructed OT displacements must be adjusted to recover the true velocities and remove RSD. These corrections require the assumption of a fiducial cosmology and bias, whose treatment will be presented in detail in Appendix~\ref{app:otderiv}.

\section{Simulated data sets}
\label{sec:simulations}
To test and validate our reconstruction and void identification algorithms, we employ two sets of $\Lambda$CDM simulations. The first is a periodic $N$-body simulation from the Aletheia suite \citep{Esposito2024} used to study the algorithm response in a controlled environment with full access to the dark matter field. The second is a realistic light-cone galaxy catalog drawn from the Buzzard mocks \citep{DeRose2019}, which includes observational effects such as RSD and realistic selection function, designed to mimic the Dark Energy Spectroscopic Instrument (DESI; \citealt{DESI}) observations. These complementary data sets allow us to assess the performance of the method both in ideal conditions and in the context of realistic survey applications.

\subsection{Aletheia}
\label{sec:aletheia}
The Aletheia simulations \citep{Esposito2024} are a suite of high-resolution $N$-body runs designed to investigate the impact of different cosmological 
growth histories on non-linear structure formation, while keeping the shape of the linear matter power spectrum, $P_\mathrm{L}(k)$, fixed. All realizations share the same linear power spectrum \(P_\mathrm{L}(k)\), determined by the physical densities \((\omega_\mathrm{b}, \omega_\mathrm{c}, \omega_\gamma)\) of baryons, cold dark matter, and photons, respectively, and by the primordial spectral index \(n_s\). They differ in the parameters governing the background evolution, namely the dark-energy density \(\omega_\mathrm{DE}\), the equation-of-state parameters \((w_0, w_a)\), and the curvature density \(\omega_K\). Each run evolves $1500^3$ dark matter particles in a periodic cubic volume of side length $1000\,h^{-1}\mathrm{Mpc}$, 
storing outputs at fixed values of the linear clustering amplitude $\sigma_{12}$ (see \citealt{Sanchez2020} for $\sigma_{12}$ definition). This ``start-stop'' integration scheme ensures full synchronization 
of particle positions and velocities at each $\sigma_{12}$ output, enabling direct comparisons across cosmologies with identical linear spectra 
but different growth histories.

In this work we employ the single realization of the baseline $\Lambda$CDM cosmology from the Aletheia suite, characterized by $\omega_\mathrm{b} = 0.02244$, $\omega_\mathrm{c} = 0.1206$, $n_s = 0.97$, $\omega_\mathrm{DE} = 0.3059$, $\omega_K = 0$, $h = 0.67$, $w_0 = -1$, $w_a = 0$, 
with $\sigma_{12}(z=0) = 0.825$ and an associated primordial normalization of $A_s = 2.127 \times 10^{-9}$. 

For our purposes, we use the $z=0$ snapshot to build 
two distinct two distinct set of unbiased, and biased tracers, respectively. The first sample consists of dark matter particles drawn from a central cubic subvolume of side length $250\,h^{-1}\mathrm{Mpc}$, with four 
random subsamples retaining 20\%, 10\%, 5\%, and 2.5\% of the particle population. The second sample consists of halos identified with the \texttt{ROCKSTAR} halo finder \citep{ROCKSTAR, Fiorilli2025}, adopting a minimum threshold of 100 particles per halo, 
corresponding to $M_\mathrm{halo} \gtrsim 2.62 \times 10^{12} M_\odot$. This selection yields $\sim 2.1 \times 10^{6}$ halos at $z=0$.

\subsection{Buzzard mocks}
\label{sec:buzzard}

The Buzzard mocks are a suite of quarter-sky (\(\sim 10{,}313\,\deg^2\)) galaxy catalogs constructed from $N$-body light-cone simulations using \texttt{Addgals} \citep{Wechsler2022}, 
an algorithm designed to reproduce the clustering properties of subhalo abundance matching models \citep{DeRose2022} in low-resolution light-cone runs. 
The underlying $N$-body simulations were performed with \texttt{L-Gadget2}, a streamlined version of \texttt{Gadget2} \citep{Springel2005} optimized for large-scale computations, 
and initialized at \(z=49\) with second-order Lagrangian perturbation theory using \texttt{2LPTIC} \citep{Crocce2006}. The linear power spectrum was generated with \texttt{CAMB} \citep{LewisBridle2002} 
assuming a $\Lambda$CDM cosmology with \(\Omega_\mathrm{m} = 0.286\), \(\Omega_\mathrm{b} = 0.046\), \(n_\mathrm{s} = 0.96\), 
\(h = 0.7\), \(\sigma_8 = 0.82\), \(\Omega_\mathrm{r} = 0\), and \(\Omega_\nu = 0\). Each light cone covers one quarter of the sky up to \(z=2.34\) and is built from 
three nested simulation boxes with \(L_\mathrm{box} = \{1050,\,2600,\,4000\}\,h^{-1}\mathrm{Mpc}\) and \(1400^3\), \(2048^3\), and \(2048^3\) particles, respectively, 
for \(z \in [0.0,0.34)\), \([0.34,0.9)\), and \([0.9,2.34)\). Each realization adopts the same cosmological parameters but varies the white-noise field used 
to generate the initial conditions.

Galaxy catalogs were generated with \texttt{Addgals}, a halo–galaxy connection model that assigns galaxies to particles in the light cone while preserving realistic clustering, 
and provides positions, velocities, and spectral energy distributions that can be integrated into broadband photometry. 
Galaxy samples mimicking the Luminous Red Galaxy (LRG) selection of the DESI \citep{Zhou2023}.

In this work, we use the real- and redshift-space LRG positions from a single realization of the Buzzard mocks.

\section{Back-in-time void finder}
\label{sec:bitvf}
We now introduce our dynamical void finder: the \texttt{Back-in-time Void Finder} (\texttt{BitVF}). \texttt{BitVF} identifies cosmic voids as regions of negative divergence in the Lagrangian displacement field reconstructed back-in-time from the observed large-scale structure through an OT approach (see Sect. \ref{sec:ottheory}). This definition is motivated by the 
physical picture of voids as regions of local mass outflow, consistent with the gravitational dynamics of structure formation \citep{SvdW2004}. 
In contrast to purely topological or density-based approaches, our method exploits the reconstructed matter dynamics to define voids in a 
physically grounded way. While previous dynamical void-finding schemes have been proposed \citep[e.g.,][]{L&W2010,Elyiv2015,MonllorBerbegal2025}, 
we present the first framework specifically designed for cosmological applications with stage IV galaxy surveys 
\citep[e.g.,][]{DESI,Euclid2025}, capable of handling complex survey geometries and the very large samples of tracers (up to tens of millions of galaxies) 
that they will deliver.

The void identification procedure is organized in three main steps:
\begin{enumerate}[itemsep=0.3em]
    \item \textbf{Displacement field reconstruction:} the displacement field $\mathbf{\Psi}(\mathbf{q})$ is reconstructed from the observed galaxy distribution using a discrete OT approach.
    \item \textbf{Divergence field computation:} the divergence of the displacement field, $\nabla_\mathbf{q}\cdot\mathbf{\Psi}$, is evaluated on a regular grid, with optional smoothing to suppress small-scale noise.
    \item \textbf{Watershed void identification:} voids are identified through a watershed segmentation of the divergence field into individual void basins from local minima of the displacement divergence, yielding dynamically motivated void regions.
\end{enumerate}
Each of these steps is described in detail in the following subsections.

\subsection{Reconstruction of the displacement field}
\label{reconstruction}
We reconstruct the Lagrangian displacement field by optimally matching each observed tracer to a random point representing its initial position. 
The goal is to recover the displacement mapping that transports the present-day tracer distribution back to a homogeneous primordial configuration. 
This reconstructed field forms the basis of our void finder, as voids can then be identified as regions undergoing net contraption when evolved backward in time.

Random catalogs containing the same number of points as the observed tracers, enabling a one-to-one assignment during reconstruction are first generated to represent unclustered Lagrangian particle distributions. In periodic simulation boxes, random particles are drawn uniformly within the volume, whereas for light-cone geometries the catalogs must reproduce the full set of observational systematic effects, as veto mask, redshift distribution \(n(z)\), and completeness variations.
 
To suppress shot noise in the random generation and obtain a stable solution, we employ multiple random realizations (typically 10–50, depending on the survey geometry), running the reconstruction independently for each tracer–random catalog pair. Each solution can be run on a single core, making the problem trivially parallelizable. The final displacement field is then defined by averaging the resulting reconstructions. 
The outcome of this procedure can be interpreted as an implicit correspondence between partitions of Lagrangian space and Eulerian positions, where the Lagrangian partition is analogous to a capacity-constrained Voronoi (or Laguerre) tessellation enforcing one tracer per cell and equal cell volumes \citep[see, e.g.,][]{Liao2018,Nikakhtar2022,Bourne2024}. In this picture, the final Lagrangian position obtained by averaging over all reconstructions coincides with the centroid of the corresponding cell.

For each tracer–random catalog pair, the tracer–random assignment is initialized with a localized seeding procedure:
\begin{itemize}[label=\textbullet,itemsep=0.3em]
    \item A tracer is randomly selected, and a sphere of radius 4 times the mean particle separation (MPS) is drawn around it.
    \item Up to 32 nearby tracers are identified in the sphere and each is matched to its nearest available random particle.
    \item Assigned tracers and their random counterparts are removed from the pool.
    \item The process is iterated until all tracers are assigned.
\end{itemize}
This initialization ensures local compactness of the displacement field and provides a suitable starting point for the optimization.

\begin{figure}
    \centering
    \includegraphics[width=0.9\linewidth]{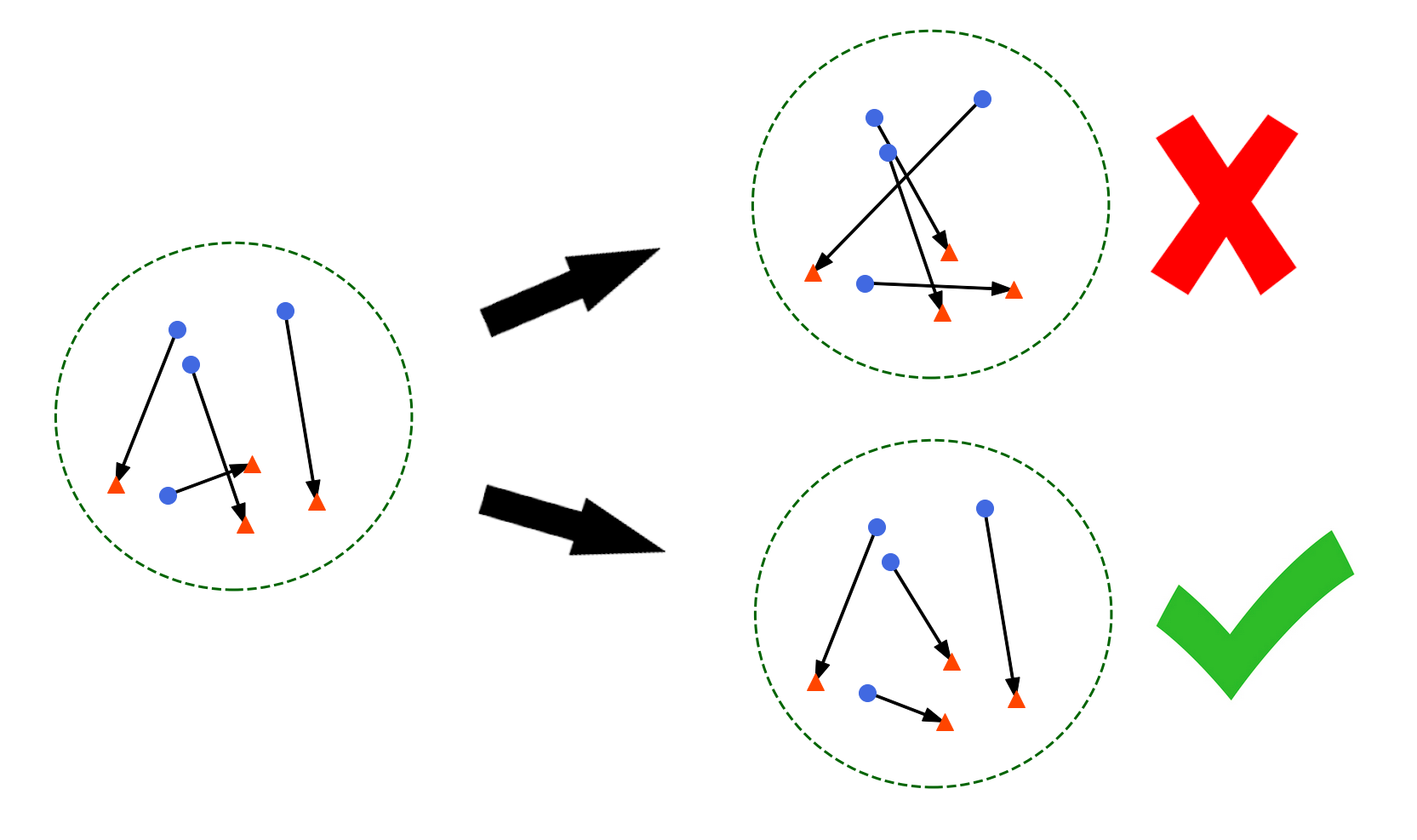}
    \caption{Two possible configurations of a quartet of displacement vectors are shown. The configuration that minimizes the total distance between 
    \(\mathbf{x}\) and \(\mathbf{q}\) positions is saved, while the other possible solutions are discarded.}
    \label{fig:swaps}
\end{figure}

The reconstruction then proceeds by minimizing the global transport cost
\begin{equation}
    \mathcal{C} = \sum_{i=1}^N \left\| \mathbf{x}_i^\mathrm{tracers} - \mathbf{x}_i^\mathrm{randoms} \right\|^2 \, ,
\end{equation}
where the sum runs over all tracer–random pairs. 
While conceptually related to the original \texttt{PIZA} algorithm \citep{piza1997}, our implementation was developed independently. Among other changes, instead of elementary pairwise swaps, we introduce a more efficient \textit{quartet-swap} scheme, which simultaneously evaluates all permutations among four local pairs, providing a favorable trade-off between accuracy and computational time. The pair-swapping optimization proceeds as follows:
\begin{itemize}[label=\textbullet,itemsep=0.3em]
    \item For each tracer, the 63 nearest neighbors are identified, restricting updates to a local region and avoiding fixed-radius biases.
    \item Quartets of tracers are randomly drawn from this neighborhood.
    \item For each quartet, all permutations of their current random assignments are tested, and the one minimizing the local \(\mathcal{C}\) is retained 
    (see Fig.~\ref{fig:swaps} for a schematic representation of the process).
    \item A full iteration corresponds to all tracers being considered once.
    \item Convergence is monitored through the fraction of successful swaps per iteration, with termination set by a threshold, $\epsilon_\mathrm{rec}$, which depends from the degree of accuracy required.
\end{itemize}

Thanks to the local structure of the optimization, the empirical scaling of the algorithm remains close to \(\mathcal{O}(N^{1.1})\). 
The implementation is parallelized with \texttt{OpenMP}, with each random realization processed independently.

To validate the accuracy of the reconstructed displacement field, we tested the reconstruction algorithm on the Aletheia halo catalog described in Sect. \ref{sec:aletheia}.  We compared the reconstructed velocities, obtained via Eq. \eqref{eq:velpsirel} and corrected for the linear bias, with the true halo velocities from the simulation. We find a Pearson correlation coefficient of $R\simeq0.77$, indicating good agreement between the reconstructed and true velocity fields. Methods and results are detailed in Appendix \ref{app:velcomp}.

\subsection{Computation of the divergence field}

Starting from the displacement field reconstructed through the OT procedure, we compute its divergence, consistently adopting the back-in-time sign convention introduced previously.

The divergence is estimated by discretizing Gauss's theorem on a three-dimensional tessellation of the survey volume. 

\subsubsection{Grid construction}

The geometry of the three-dimensional grid on which we compute the divergence field is determined by the tracer catalog, with resolution chosen to balance accuracy and efficiency. Cells must be small enough to resolve local features of the displacement field while ensuring computational feasibility over the full survey volume.

\paragraph{Periodic boxes.}
For periodic simulation boxes, we adopt a Cartesian grid of cubic cells (\textit{voxels}) aligned with the box axes. The cell size is set to
\begin{equation}
    l_\mathrm{cell} = k \times \mathrm{MPS} \, ,
\end{equation}
where $\mathrm{MPS} = (V/N)^{1/3}$, with $V$ and $N$ denoting the box volume and the total number of objects, respectively, is the mean particle separation of the catalog, and $k$ is a tunable parameter ($0.5 \lesssim k \lesssim 1$; see Appendix~\ref{app:impact} for a detailed discussion).
A buffer region around the box ensures that all tracer–random segments remain fully contained.

\paragraph{Light-cone surveys.}
For realistic survey geometries, we construct a curvilinear, adaptive grid centered on the observer. Cells are defined as intersections of three families of boundaries:
\begin{itemize}[label=\textbullet,itemsep=0.1em]
  \item \textbf{Radial shells} at comoving distances \(\{r_i\}\), with depth
  \begin{equation}
    \Delta r_i = k \times \mathrm{MPS}(z_i) ,
  \end{equation}
  where \(\mathrm{MPS}(z_i)\) is the mean tracer separation computed at the redshift $z_i$ corresponding to the shell midpoint.
  \item \textbf{Declination cones} with angular width chosen so that the physical height of the cell at \(r_i\) is \(\sim k \times \mathrm{MPS}(z_i)\).
  \item \textbf{Right-ascension wedges} adjusted analogously to preserve the transverse physical size.
\end{itemize}
Cells near the poles are closed with cap-like geometries to avoid wedge collapse. The resulting mesh spans the full comoving survey volume, conforming to the angular mask and veto regions, while maintaining continuous, non-overlapping cell boundaries. By construction, the effective physical resolution adapts with redshift to track the local MPS, thereby suppressing redshift-dependent biases in void sizes and abundances. At sufficiently large comoving distances, cells asymptotically approach a cubic geometry.

\begin{figure}
    \centering
    \includegraphics[width=1.0\linewidth]{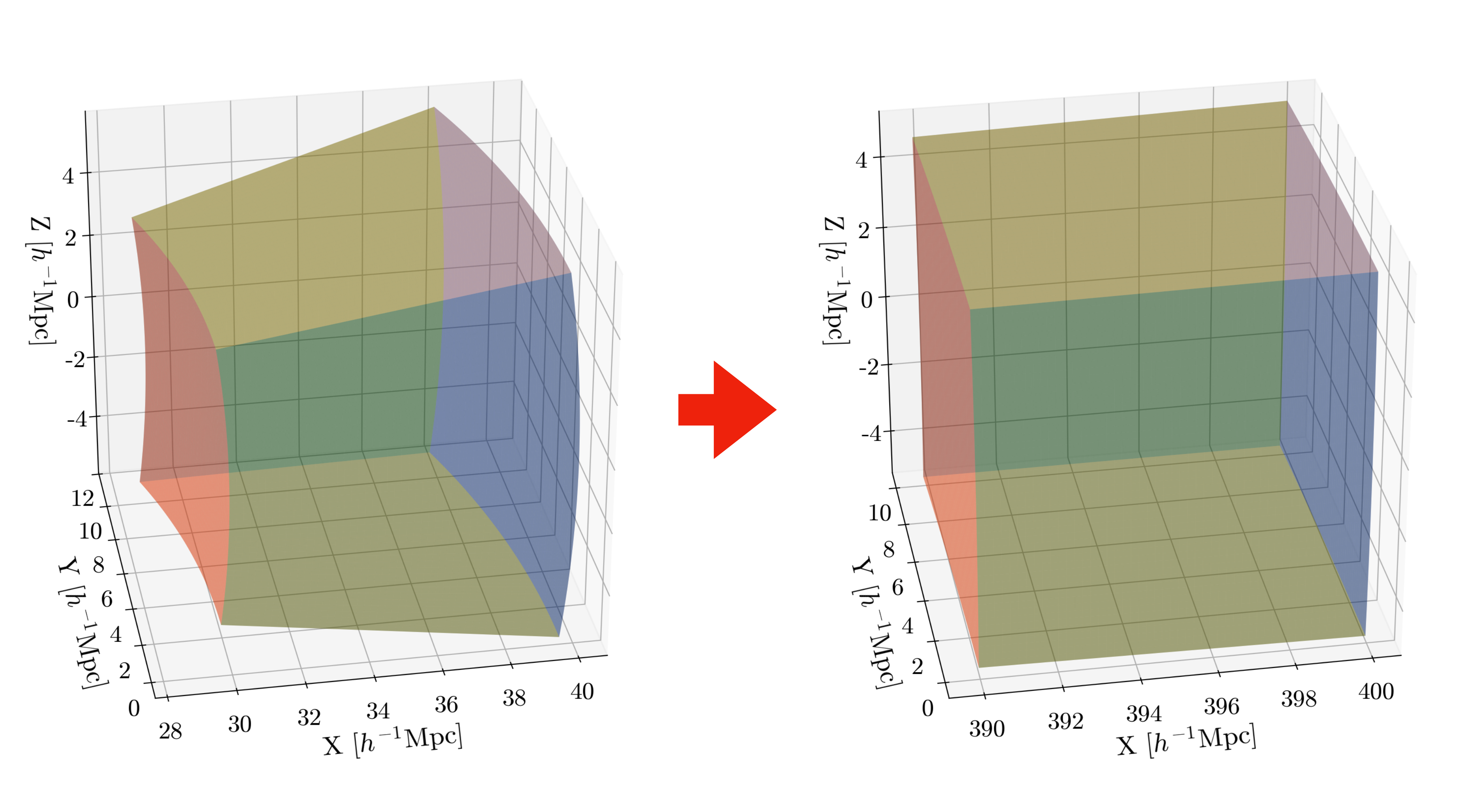}
    \caption{\textit{Left}: Example of a light-cone cell, defined by the intersection of two concentric spheres, two meridional planes, and two cones whose vertices lie at the center of the Cartesian coordinate system and whose axes coincide with the polar axis.  
    \textit{Right}: The same cell located at a large comoving distance from the observer, where its geometry becomes approximately cubic.
}
    \label{fig:cell}
\end{figure}

Examples of the resulting cell shape and behavior at different distances from the origin are shown in Fig. \ref{fig:cell}.

\subsubsection{Divergence estimation}

For each cell \(c\) with volume \(V(c)\) and for each random realization \(r = 1,\dots,N_{\mathrm{rec}}\), we identify the set \(\mathcal{S}_c^{(r)}\) of displacement vectors that cross the faces of \(c\). Each intersection contributes a flux proportional to the projection of the displacement vector onto the outward normal of the crossed face, weighted by the intersected surface area and normalized by the cell volume. Averaging over random realizations suppresses stochastic fluctuations and enforces consistency with the survey selection function.

The divergence in cell \(c\), noted as $\Theta(c)$, is thus computed as
\begin{equation}
\Theta(c) = (\nabla \cdot \boldsymbol{\Psi})_c
= \frac{1}{V(c)\, w(c)}
\sum_{i \in \mathcal{S}_c}
(\boldsymbol{\Psi}_i \cdot \hat{n}_i)\, \Delta S_i \, ,
\end{equation}
where the sum runs over all intersections contributing to cell \(c\). $\boldsymbol{\Psi}_i$ denotes the displacement vector, $\hat{\mathbf{n}}_i$ the outward normal at the intersection, and $\Delta S_i$ the intersected face area. The normalization factor 
\begin{equation}
   w(c) = \frac{1}{N_{\mathrm{rec}}}
\sum_{r=1}^{N_{\mathrm{rec}}}
\sum_{i \in \mathcal{S}_c^{(r)}} 1 
\end{equation}
accounts for the number of intersections contributing to the cell across all realizations. 

Although the displacement field is formally defined in Lagrangian space, the divergence is computed on a discrete Eulerian grid, which provides an accurate approximation on scales larger than the typical shell-crossing length.
 
\subsubsection{Interpolation and smoothing}

In cases with a limited number of random realizations, some grid cells may receive no intersecting displacement segments, leaving them empty and potentially introducing unphysical artifacts. 
To produce a continuous divergence field, $\Theta_\mathrm{interp}$, suitable for the following watershed segmentation, we perform a conservative local interpolation. 
For each empty cell \(c\), we identify a neighborhood \(\mathcal{N}\) of nearby cells within a search radius \(R_\mathrm{fill}\) (typically 2-3 cell units) that contain valid divergence estimates. 
The interpolated value is then obtained through an inverse-distance weighted average,
\begin{equation}
  \Theta_\mathrm{interp}(c) =
  \frac{\sum_{c' \in \mathcal{N}} \omega(c') \, \Theta(c')}
       {\sum_{c' \in \mathcal{N}} \omega(c')} ,
  \qquad \omega(c') = \frac{1}{d(c,c')^\gamma}\,,
\end{equation}
where $\omega(c')$ is the weight, \(d(c,c')\) is the distance between cell centers and \(\gamma \ge 1\) controls the distance weighting (we adopt \(\gamma=1\) by default). 
Cells with no valid neighbors within \(R_\mathrm{fill}\) are flagged and excluded from the segmentation; this occurs only in small, heavily masked regions. 
This scheme preserves the local structure of the field while leaving already sampled cells unchanged.

To further suppress small-scale noise and enhance coherent, large-scale outflows, the interpolated divergence field $\Theta_\mathrm{interp}$ can be optionally smoothed with a three-dimensional Gaussian kernel. The resulting smoothed divergence field, denoted as $\Theta_\mathrm{sm}$, is given by
\begin{equation}
\label{eq:gauss_smooth}
\Theta_\mathrm{sm}(c) =
\frac{\displaystyle\sum_{c'} \Theta_\mathrm{interp}(c') \, \exp\Bigl[-\frac{d(c,c')^2}{2\sigma_\mathrm{sm}^2}\Bigr]}
{\displaystyle\sum_{c'} \exp\Bigl[-\frac{d(c,c')^2}{2\sigma_\mathrm{sm}^2}\Bigr]} \,,
\end{equation}
where $\sigma_\mathrm{sm}$ is expressed either in units of the local MPS or in absolute comoving units. In practice, choosing $\sigma_\mathrm{sm} \ll 1\,\mathrm{MPS}$ preserves fine structure, thereby maximizing the number of identified void centers and resolving sub-void features, while values $\sigma_\mathrm{sm} \sim 1\,\mathrm{MPS}$ effectively suppress sub-cell fluctuations, improving the stability of cross-correlations and facilitating comparisons with analytic void statistics.

A top-hat filter of radius \(R = \sigma_\mathrm{sm}\) can be used alternatively if desired. 
Note that any smoothing modifies the effective catalog completeness for small voids; all smoothing choices used in the analysis are explicitly reported in the results and their impact quantified.

Applying smoothing after interpolation ensures that no empty cells remain, producing a continuous, gap-free divergence field suitable for the subsequent watershed subdivision. 

\subsection{Watershed-based void identification}

Watershed algorithms are widely used for cosmic void identification in both geometric and density-based finders \citep{Platen2007,Zobov2008,L&W2010,Elyiv2015,Sutter2015,Revolver2019}. In these methods, the watershed transform is applied to the density field to delineate underdense regions. Here, the same technique is applied to the smoothed divergence of the reconstructed back-in-time OT displacement field. This formulation identifies voids dynamically rather than topologically, shifting the focus from local density to global dynamical information.

\begin{figure}
\centering
\includegraphics[width=1.0\linewidth]{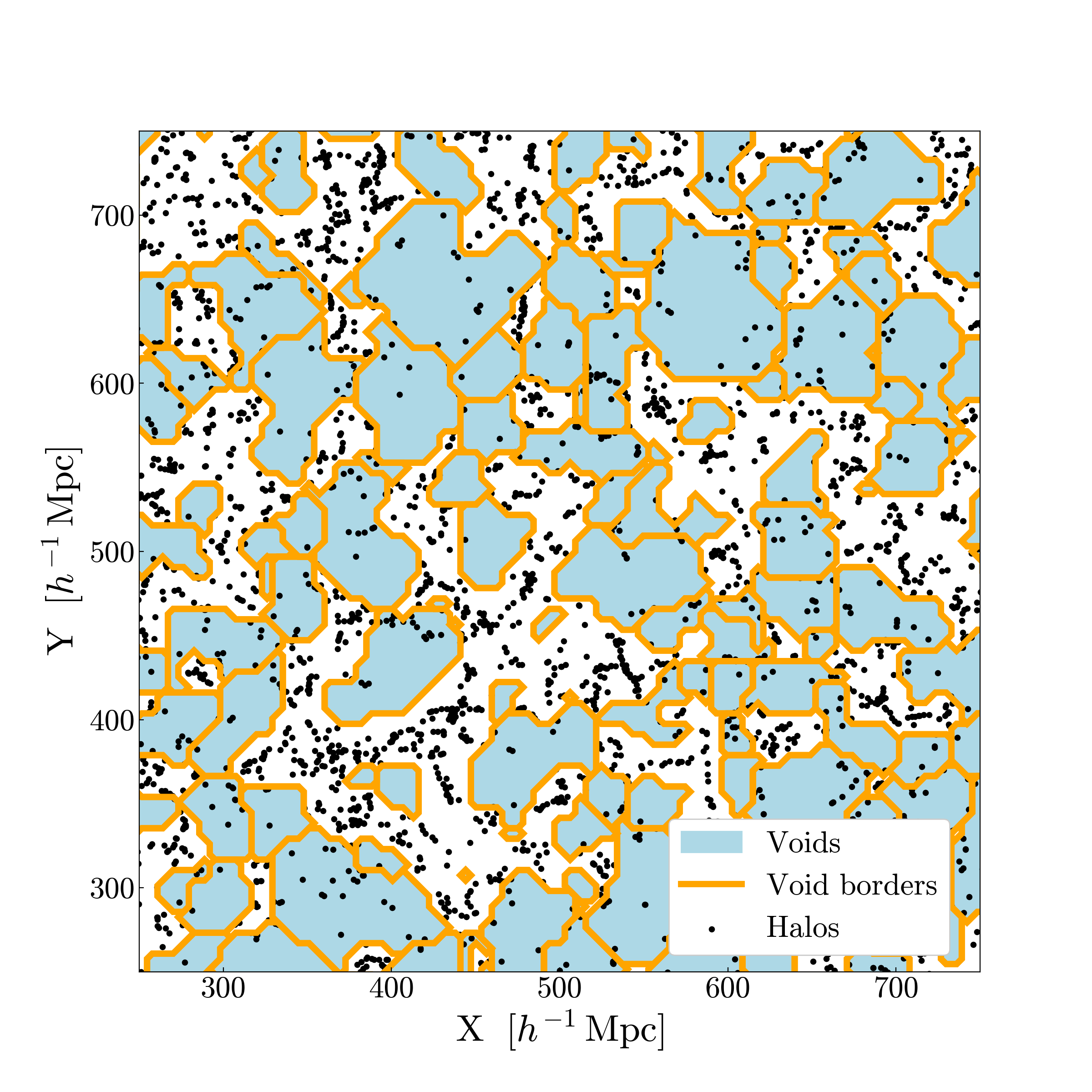}
\caption{A $5 \,h^{-1}\mathrm{Mpc}$-thick slice extracted from the $500^3 \,h^{-3}\mathrm{Mpc}^3$ core of the Aletheia halo sample.
The figure shows the resulting watershed segmentation of the divergence field, with voids shown in light blue and their boundaries in orange. Regions of positive divergence (overdensities) are shown in white, while halos are shown in black.}
\label{fig:voidBorders}
\end{figure}

Starting from the smoothed divergence field, voids are identified by associating each unflagged cell with negative divergence to a local minimum of the divergence field. The assignment proceeds iteratively: for each cell, all adjacent neighbors are examined, and the cell is moved to the neighbor with the most negative divergence, repeating the process until a local minimum is reached.
The set of all cells converging to the same minimum defines a connected basin, corresponding to a void. Each void is assigned an effective radius
\begin{equation}
R_\mathrm{eff} = \left(\frac{3}{4\pi} \sum_{c\in\mathrm{basin}} V(c) \right)^{1/3} \, ,
\end{equation}
where the sum extends over all cells in the basin and $V(c)$ is the volume of each cell.

For each void, the center is initially defined as the position of the local minimum $\mathbf{x}_\mathrm{min}$ of the divergence field associated with the corresponding basin. This estimate is then refined by applying a weighted correction based on the divergence values of the neighboring cells. Specifically, the void center coordinates $\mathbf{x}_\mathrm{c}$ are fine-tuned according to
\begin{equation}
\mathbf{x}_\mathrm{c} = \mathbf{x}_\mathrm{min} + 
\frac{\sum_{c'} \Theta(c')\, (\mathbf{x}_{c'} - \mathbf{x}_\mathrm{min})}{\sum_{c'} |\Theta(c')|} \, ,
\end{equation}
where the sum runs over the 26 neighboring cells $c'$ surrounding the minimum, and $\Theta(c')$ is the divergence of the reconstructed displacement field. Here, $\mathbf{x}_{c'} - \mathbf{x}_\mathrm{min}$ is the vector from the minimum to the center of cell $c'$, and the resulting displacement is naturally bounded by the extent of the neighboring cells, ensuring that the refined center remains confined within the watershed basin while shifting toward the region of strongest coherent outflow.

This watershed method partitions the negative-divergence domain into separate basins centered on local minima, naturally splitting extended underdensities into smaller, non-overlapping substructures tracing regions of coherent outflow in the reconstructed displacement field. An example of the resulting watershed segmentation is shown in Fig.~\ref{fig:voidBorders}. 

The resulting void size function therefore differs from those based on fixed density thresholds, which predict the abundance of spherical regions of given mean density \citep{SvdW2004, vdn2013, Verza2024}.

\subsection{Performance benchmarking}

\begin{table*}
    \centering
    \captionof{table}{Running times}
    \begin{tabular}{cccccccc}
        \specialrule{.15em}{.2em}{0.2em} 
	    \specialrule{.05em}{.05em}{0.3em} 
        \textit{Fraction} & $t_\mathrm{rand} \; [s]$ & $t_\mathrm{rec}\; [s]$ & $t_\mathrm{disp}\; [s]$ & $t_\mathrm{div}\; [s]$ & $t_\mathrm{water}\; [s]$ & $t_\mathrm{tot}\; [s]$ & $\overline{{N}_\mathrm{v}}$ \\[0.3em]
        \specialrule{.1em}{.2em}{0.8em} 
        1/100 & $0.17\pm0.02$ & $3.07\pm0.18$ & $0.14 \pm 0.04$ & $0.62\pm0.16$ & $0.16\pm0.04$ & $4.32 \pm 0.30$ & $90.80\pm3.18$ \\[0.5em]
        1/10 & $1.17\pm 0.08$ & $40.64\pm 0.63$ & $1.33\pm 0.24$ & $5.51\pm0.66$ & $1.46\pm 0.24$ & $51.02\pm0.92$ & $855.75\pm 8.60$ \\[0.5em]
        1 & $8.09\pm 0.56$ & $585.73\pm20.25$ & $14.05\pm 1.32$ & $51.68\pm3.85$ & $15.32\pm 1.37$ & $686.74\pm21.45$ & $8910.30\pm 24.25$\\[0.5em]
        \specialrule{.1em}{.2em}{0.8em} 
    \end{tabular}
    \label{times}
\tablefoot{The table shows the average running times for each relevant part of the algorithm, the total running time, and the average number of voids, along with their associated standard deviations. The errors correspond to the standard deviation across the 100 independent realizations. The data are presented for the 1\%, 10\%, and full halo samples.}
\end{table*}

To evaluate the computational performance and statistical robustness of our void finder, we applied the algorithm to the Aletheia halo catalog described in Sect.~\ref{sec:aletheia}, performing 100 independent runs. Three different tracer numbers were tested by selecting nested sub-volumes around the center of the simulation box, containing the full sample, 10\%, and 1\% of the total halos (\(\sim\!2.1 \times 10^{6}\) in the full volume). This procedure preserves the local halo number density, effectively maintaining the void resolution and reconstruction response across volumes. By fixing the local halo number density, this setup allows us to characterize the computational scaling of the method independently of other sampling or cosmological effects.

All runs were executed on a single simulation realization using 50 CPU cores on an Intel\textregistered\, Xeon\textregistered\, E5-2680 v4 @ 2.40 GHz processor. 
We adopted \(N_\mathrm{rec}\!=\!50\), \(\epsilon_\mathrm{rec}\!=\!10^{-3}\), \(l_\mathrm{cell}\!=\!1/\sqrt[3]{2}\,\mathrm{MPS}\), and \(\sigma_\mathrm{sm}\!=\!1\,\mathrm{MPS}\). 
To assess computational scaling, internal checkpoints recorded the runtime of each major stage of the pipeline, enabling a detailed performance analysis as a function of input size.

Table~\ref{times} reports the mean runtimes (in seconds) and standard deviations for each pipeline stage: random catalog generation 
(\(t_\mathrm{rand}\)), OT reconstruction (\(t_\mathrm{rec}\)), displacement field construction (\(t_\mathrm{disp}\)), divergence computation 
(\(t_\mathrm{div}\)), and watershed segmentation (\(t_\mathrm{water}\)). The total runtime (\(t_\mathrm{tot}\)) and the average number of identified voids 
(\(\overline{N}_\mathrm{v}\)) are also listed.

\begin{figure}
    \centering
    \includegraphics[width=1.\linewidth]{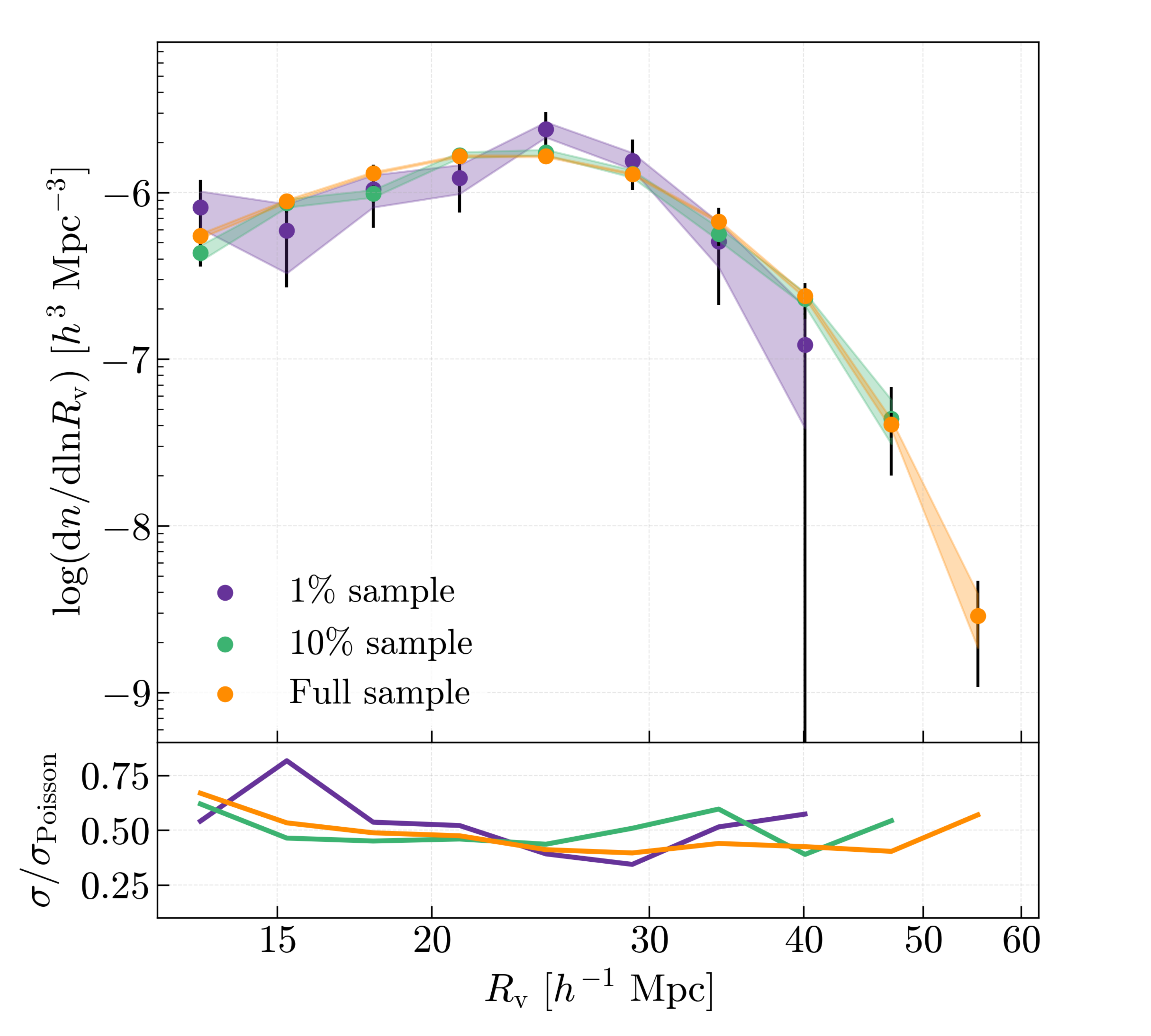}
    \caption{\textit{Top}: VSFs for the 1\%, 10\%, and full halo samples, shown in purple, green, and orange, respectively. The errorbars represent the associated Poissonian error, while the shaded bands indicate the standard deviation for the void counts across the 100 runs, for each sample. \textit{Bottom}: The ratios between the intrinsic stochastic error of the method and the Poissonian error.
}
    \label{fig:scatter}
\end{figure}

\begin{figure*}
    \centering
    \includegraphics[width=1.\linewidth]{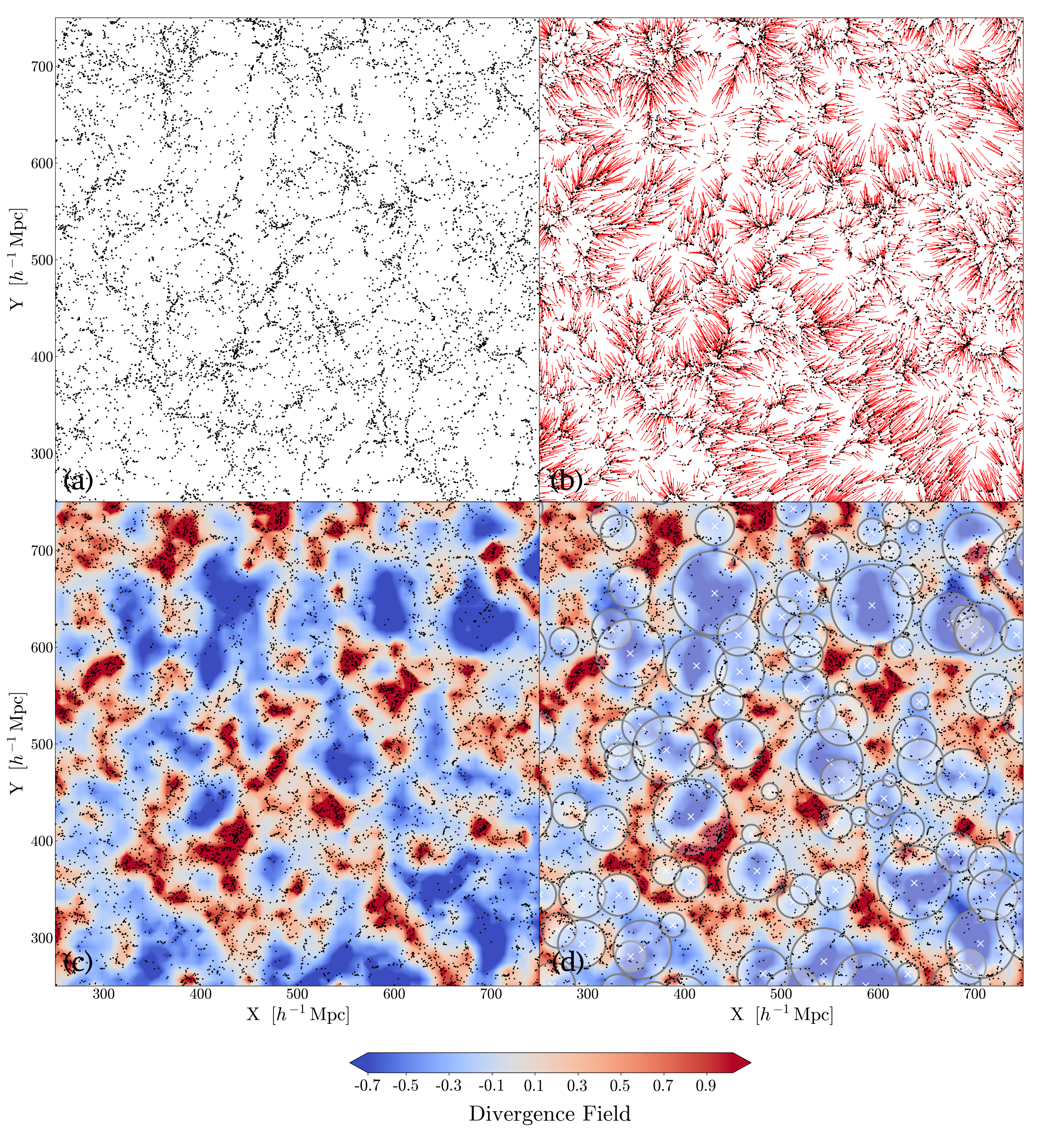}
    \caption{A slice of $20 \,h^{-1}\mathrm{Mpc}$ thickness from the $500^3 \,h^{-3}\mathrm{Mpc}^3$ core of the halo sample from the Aletheia simulations. Different stages of the void identification procedure are shown. (\textit{a}) The Eulerian positions of the halos at $z\!=\!0$. (\textit{b}) The reconstructed velocities of the halos. (\textit{c}) The divergence field superimposed on the halo positions. Blue regions indicate zones with negative divergence of the back-in-time displacement field, corresponding to regions from which halos tend to escape when evolved forward in time. Red regions indicate positive divergence, marking regions of local mass inflow. (\textit{d}) The same as panel c, with the identified voids overplotted. Voids are identified using the watershed algorithm introduced above and are represented as circles for graphical clarity.
 }
    \label{fig:bitVF}
\end{figure*}
We find that the reconstruction phase dominates the total runtime, with its cost scaling according to the number of tracers and the complexity of the transport map. 
As expected, all stages exhibit increased runtimes for higher halo numbers. The overall scaling of the total runtime is approximately linear, with a power-law fit 
across the three sampling levels yielding \(t \propto N^{1.1}\), where \(N\) is the number of tracers. This quasi-linear behavior ensures that the algorithm remains computationally tractable even at full resolution. Furthermore, providing a precomputed displacement field, potentially obtained from an external reconstruction, substantially reduces the total runtime by bypassing the most computationally demanding stage.

Concurrently, we computed the VSF for each run to quantify the stochastic scatter inherent to the method. We stress that this VSF is \textit{uncleaned}, in the sense that the identified voids are not rescaled or adjusted to match any theoretical model. A comprehensive exploration of the method response to internal parameters, including smoothing scale, cell resolution, number of reconstruction iterations, and stopping threshold, is presented in Appendix \ref{app:impact}.

Figure~\ref{fig:scatter} demonstrates that the VSF remains stable across the different tracer numbers, as expected given that the local halo number density was preserved across the nested volumes, and despite the increasing relevance of sampling variance in the smaller volumes.
As shown in the lower panel, the stochastic fluctuations introduced by the algorithm are systematically smaller than Poisson uncertainties, confirming the robustness of the method against reduced tracer counts and the negligible impact of random catalog initialization on large-scale void statistics. These stochastic contributions can be included in the total error budget when comparing the VSF with theoretical predictions or independent measurements, and can also provide a reliable estimate of the overall uncertainty in repeated reconstructions of the same sample using \texttt{BitVF}. 

\section{Method tests}
\label{sec:tests}

In this section, we validate our void-finding algorithm using the Aletheia simulation, aiming to quantify its performance in an idealized, controlled environment that allows a direct assessment of both the displacement reconstruction quality and the accuracy in void identification, free from observational uncertainties.

The first validation focuses on the halo catalog at $z=0$, with void identification performed using the parameter configuration $N_{\mathrm{rec}} = 50$, $\epsilon_\mathrm{rec} = 10^{-3}$, $\sigma_{\mathrm{sm}} = 1,\mathrm{MPS}$, and $l_\mathrm{cell}=1/\sqrt[3]{2},\mathrm{MPS}$. This choice maximizes the purity of the void catalog by requiring a highly converged reconstruction and suppressing small-scale noise through smoothing, thereby emphasizing structures driven by coherent large-scale flows.

Figure~\ref{fig:bitVF} illustrates the main stages of the detection procedure for a $20\,h^{-1}\mathrm{Mpc}$-thick slice through the central region of the simulation box. Panel (a) displays the halo distribution in Eulerian space. Panel (b) presents the result of the back-in-time OT reconstruction, where halo displacements clearly point from overdense regions toward surrounding underdensities, showing the consistency of the inferred mass flows. Panel (c) shows the resulting smoothed divergence field, which traces the large-scale structure of the halo distribution and highlights the coherence of the velocity field. Panel (d) displays the voids identified by \texttt{BitVF}, approximated as spheres for visualization. The apparent overlaps result from this spherical approximation and the finite thickness of the slice; only voids whose centers are within the slice are shown.

\subsection{Main void statistics}

We now compare, after applying a border cut of $30 \, h^{-1}\mathrm{Mpc}$, the VSF obtained with our method to that produced by the modified \texttt{ZOBOV} \citep{Zobov2008} algorithm included in \texttt{REVOLVER} \citep{Revolver2019}, which identifies voids starting from a Voronoi tessellation of the tracer distribution. The same border cut is applied to both \texttt{BitVF} and \texttt{REVOLVER} void catalogs to remove edge voids, which are typically incomplete and therefore unreliable. The chosen value represents a compromise between the size of the largest voids and the need to retain sufficient statistics. In \texttt{REVOLVER}, the volume of each Voronoi cell is then used to estimate a local density field, with underdense regions corresponding to large cells. Local minima in this density field define the seeds of watershed basins, which are then grown by joining neighboring cells until the ridge between adjacent basins exceeds a predefined density threshold. These basins are merged into voids, aiming to cover the totality of the volume. The center of each void is defined topologically as the circumcenter of the tetrahedron formed by the four most underdense and mutually adjacent Voronoi cells within the void, providing a robust estimate for the location of the most underdense region.

 \begin{figure}
    \centering
    \includegraphics[width=1\linewidth]{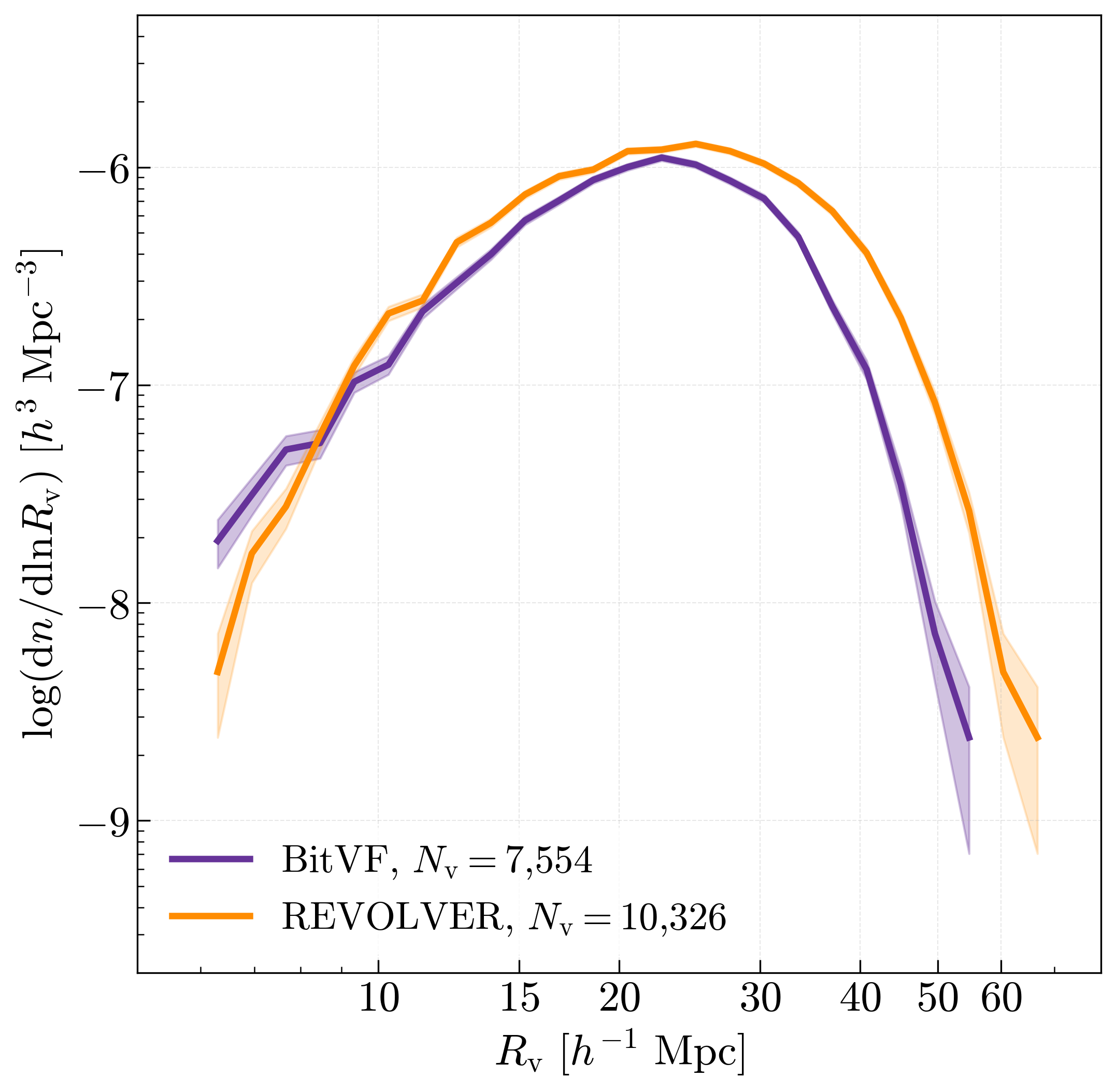}
    \caption{VSFs of voids identified in the halo sample of the Aletheia simulations, shown in orange for \texttt{REVOLVER} and in purple for \texttt{BitVF}. The associated uncertainties, shown as colored bands, are assumed to be Poissonian.}
    \label{fig:vsf}
\end{figure}

\begin{figure}
    \centering
    \includegraphics[width=1\linewidth]{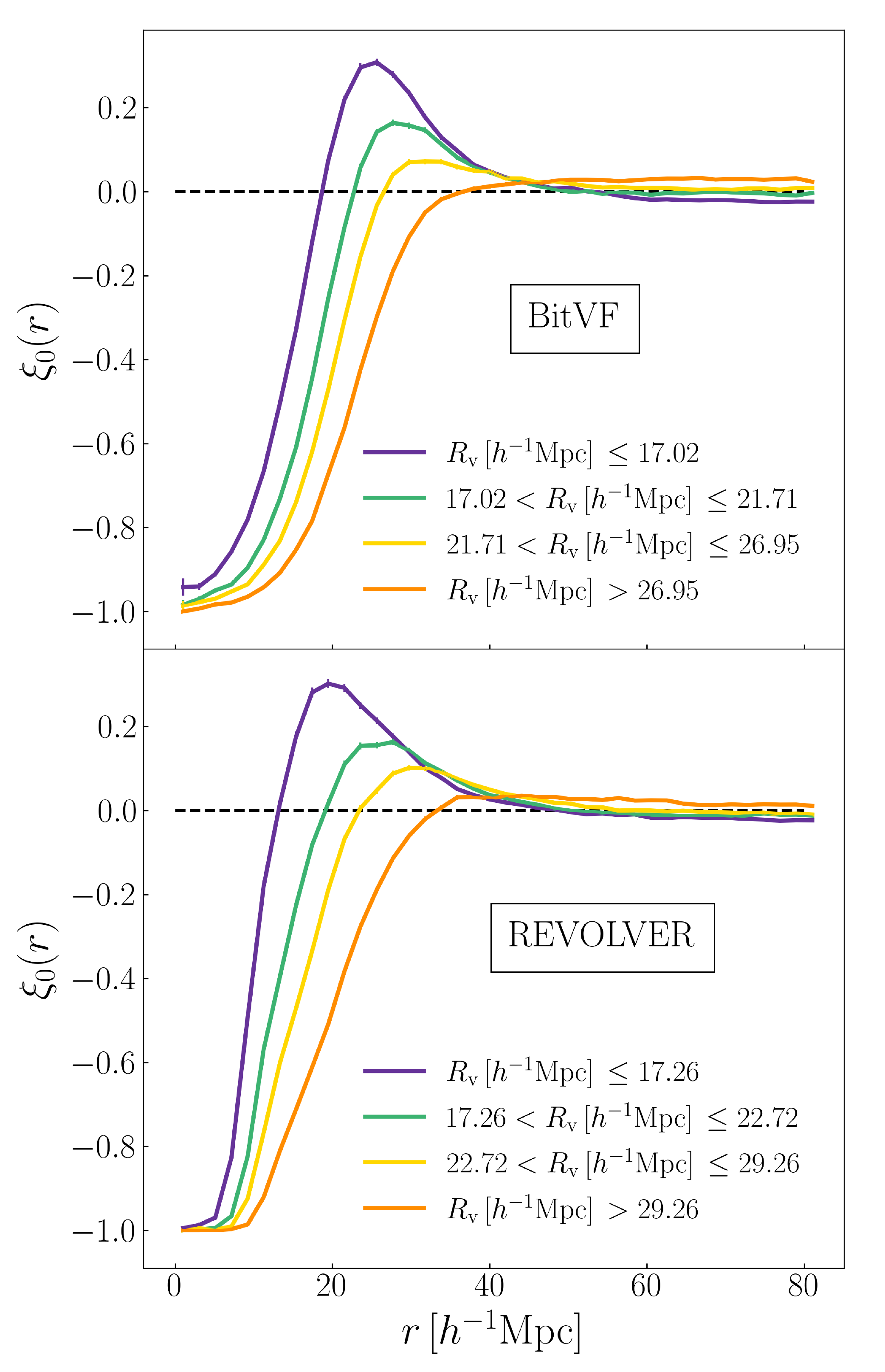}
    \caption{VGCF monopoles from the voids identified in the halo sample of the Aletheia simulations. The void samples are subdivided into four bins in size, corresponding to the ranges defined by the 0-25th, 25-50th, 50-75th, and 75-100th percentiles of the radius distribution, starting from the largest voids. \textit{Top} and \textit{bottom} panels show the results for voids identified in \texttt{BitVF} and \texttt{REVOLVER}, respectively. Errorbars are taken from the diagonal of the associated covariance matrices.}
    \label{fig:CC}
\end{figure}

The comparison of the two VSFs from \texttt{BitVF} and \texttt{REVOLVER} is shown in Fig.~\ref{fig:vsf}. Both distributions exhibit a similar overall shape, demonstrating consistency between the methods. However, the VSF from \texttt{BitVF} is systematically shifted toward smaller radii. This behavior is expected: by construction, \texttt{BitVF} grows void basins only within regions of negative divergence. As a result, for a given underlying structure, the expansion stops at the boundary where the divergence changes sign, excluding the compensating overdense shells. Consequently, the resulting voids have on average smaller effective radii compared to those identified by density-based watershed schemes that extend over the full field.

As discussed in Appendix \ref{app:impact}, the uncleaned VSF depends mainly on the parameters adopted for the divergence computation, $\sigma_\mathrm{rm}$ and $l_\mathrm{cell}$. The reference configuration used here was chosen to mimic the behavior of Voronoi-based finders such as \texttt{REVOLVER}, which resolution is intrinsically given by the local MPS of the tracers.

We further compare the two void catalogs by analyzing the monopole of the void–galaxy correlation function (VGCF), shown in Fig.~\ref{fig:CC}, which corresponds to the stacked void density profiles obtained by averaging over the full void sample at fixed physical separations, without rescaling, in this case, distances by the individual void radii. To minimize possible binning mismatches between the two methods, we subdivide our void catalogs into four radius bins defined by the 0–25th, 25–50th, 50–75th, and 75–100th percentiles of the radius distribution, starting from the largest voids, which are more likely to be jointly identified by both finders. 

The overall shape of the monopole is remarkably similar between the two catalogs and consistent with expectations from the literature (e.g. \citealt{Hamaus2014, Schuster2023}): larger voids display broader, shallower profiles with mild compensation walls, while smaller ones exhibit steeper slopes and sharper overdense ridges. Profiles from \texttt{BitVF} appear smoother, whereas those from \texttt{REVOLVER} retain a more top-hat–like structure. This difference likely arises from the dynamical nature of \texttt{BitVF}, which not only centers voids in the minimum density regions (corresponding to the points of maximum divergence) but also accounts for coherent mass flows, leading to smoother transitions between void interiors and boundaries. Such behavior may also prove a reduced sensitivity of \texttt{BitVF} to spurious, poorly defined voids.

\subsection{Dark matter sample and subsampling response}
A void-finding algorithm that is robust to variations in tracer density (i.e., tracer sparseness) is essential to ensure that void properties remain consistent across different resolutions. Such flexibility allows the method to adapt to surveys with varying tracer populations or redshift-dependent number densities, ensuring that increasing resolution improves the identification of small-scale structures without biasing the statistics of large voids.

\begin{figure*}
    \centering
    \includegraphics[width=1.\linewidth]{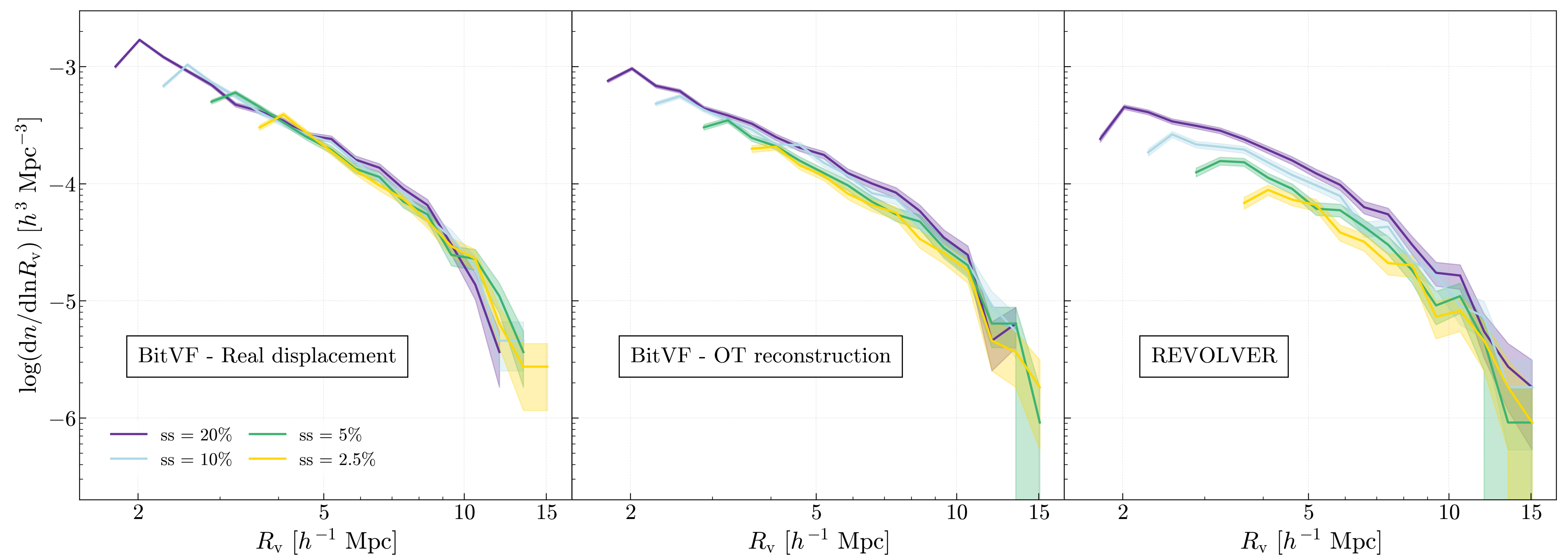}
    \caption{Cleaned VSFs from the voids identified in the subsampled dark matter catalogs of the $250^{3}\,h^{-3}\mathrm{Mpc}^3$ central region of the Aletheia simulations. \textit{Left}: VSFs from \texttt{BitVF} voids, identified using the displacement field reconstructed from the real particle velocities, 
    serving as the reference. \textit{Center}: VSFs from \texttt{BitVF} voids, identified using the displacement field obtained through our OT
    reconstruction algorithm. \textit{Right}: VSFs from \texttt{REVOLVER} voids, for comparison with a topological method. Errors (colored bands) are assumed to be Poissonian for all the measurements.}
    \label{fig:ss}  
\end{figure*}

We assess the performance of our method for different sparseness levels, testing it on four density subsamples of the Aletheia dark matter distribution, corresponding to 20\%, 10\%, 5\%, and 2.5\% of the total particle population. 
These subsamples are drawn from the central $250\,h^{-1}\mathrm{Mpc}$ region of the full simulation box (see Sect.~\ref{sec:aletheia}). 
This controlled setup enables us to probe two main aspects of the methodology. 
First, by randomly subsampling the tracers, we test how the OT reconstruction responds to reduced resolution in highly non-linear regions. Uniform subsampling sparsely samples dense structures such as halos, reducing central density spikes and thereby mitigating the impact of shell crossing on the reconstructed displacement field and void identification. 
Second, it allows us to assess the stability of the void finder under different tracer sparseness levels.

For this analysis, we adopt the \textit{cleaned} VSF (see \citealt{Ronconi2017}) as the primary probe. 
Each void is rescaled to a sphere enclosing a mean density contrast of $\delta = -0.8$, and overlaps are resolved by retaining only the largest void within each group. 
This cleaning scheme differs from the one proposed by \citet{vdn2013}, who based overlap removal on internal densities without a rigorous theoretical justification, and is more closely aligned with the approach of \citet{Verza2024}, which maximizes void volumes while preserving the largest structure in each overlap. 
A border cut of $20\,h^{-1}\mathrm{Mpc}$ is applied to mitigate edge effects, a slightly smaller value than the $30\,h^{-1}\mathrm{Mpc}$ used previously, reflecting the higher tracer density of the samples.

The cleaned VSF provides a particularly suitable test for this, as it is expected to yield a stable distribution at large radii regardless of sampling. Smaller voids, in contrast, are expected to appear progressively with increasing tracer density, reflecting the improved resolution of small-scale underdensities, with the effective sensitivity determined by the MPS.

Figure~\ref{fig:ss} summarizes the results of the test, showing three sets of cleaned VSFs. The first set corresponds to voids identified by running \texttt{BitVF} directly on the approximated linear back-in-time displacement field computed via Eq. \eqref{eq:velpsirel} from the real, non-linear, particle velocities in the simulation. These catalogs serve as a reference for our stability test. The second set shows the VSFs obtained with \texttt{BitVF} when the displacement field is reconstructed using our discrete OT algorithm, with $N_\mathrm{rec}=50$ random realizations and a stopping threshold $\epsilon_\mathrm{rec}=10^{-3}$. In both cases, no Gaussian smoothing is applied to retain sensitivity to the smallest voids. The third set presents the VSFs identified with the \texttt{ZOBOV}-based \texttt{REVOLVER} algorithm, providing a benchmark to evaluate the relative efficiency and stability of our method.

The comparison reveals several key trends. The reference VSFs obtained from the true linear displacement field are remarkably stable across the different subsamples. As expected, the cut-off in the number of voids shifts toward smaller radii as the sampling density increases, indicating that finer structures become accessible with denser tracer populations. Importantly, the overall shape of the VSF remains stable, showing that voids identified at large scales are robust to significant reductions in tracer density.

VSFs obtained from the OT reconstructed displacement field remain generally compatible across subsamples, especially at large radii. Differences appear primarily between the sparsest sample and the others, particularly at intermediate and small scales, whereas the VSFs from the denser subsamples are mutually consistent. This indicates that the observed deviations reflect the intrinsic difficulty of preserving the internal structure of voids at low resolution, rather than high tracer density or associated non-linearities, limitations associated with finite sampling and resolution rather than the algorithm itself.
In contrast, the cleaned VSFs produced with \texttt{REVOLVER} show a stronger dependence on tracer density. Both the amplitude and shape of the distributions vary significantly across subsamples, especially at small and intermediate radii. This instability reflects the higher sensitivity of topological void finders to sampling density: sparse samples amplify discreteness effects in the Voronoi tessellation, producing spurious structures and unstable statistics.

The general consistency across subsamples, shown by the cleaned \texttt{BitVF} VSFs, demonstrates that the original large-scale minima identified by \texttt{BitVF} remain robust, effectively preserved despite the fragmentation induced by significant changes in resolution and sparseness level. Our method can handle tracer densities down to $\sim 0.5\text{-}1\,h^3\,\mathrm{Mpc}^{-3}$ without significant degradation. This confirms the robustness of the reconstruction procedure both in dark matter simulations and in practical applications, given that the tracer densities of current spectroscopic and photometric surveys are considerably lower than those used in these tests: the MPS in spectroscopic surveys such as 4MOST and DESI is typically $5\text{-}10\,h^{-1}\mathrm{Mpc}$ \citep{Verdier2025, DESI}, while deep photometric surveys like \textit{Euclid} and the Large Synoptic Survey Telescope (LSST) achieve higher tracer densities with mean separations of $\sim 1\text{-}3\,h^{-1}\mathrm{Mpc}$ \citep{EuclidWIDE, LSST2}.  
Given this range of tracer densities, dynamical void finders such as \texttt{BitVF} are particularly well suited for upcoming surveys, as they can reliably identify true underdensities and maintain stable void statistics, in contrast to purely topological methods whose results may depend more strongly on tracer sparsity and resolution.

\section{Light-cone applications}
\label{sec:lightcone}

While the previous analysis focused on idealized cubic volumes in real space, real observations are performed in redshift space and within a light-cone geometry, where geometrical and dynamical distortions become unavoidable.
Voids identified in redshift space encode valuable cosmological information through distortions induced by RSD and the AP effect, which can be probed, for example, through the VGCF (e.g., \citealt{Hamaus2015,Hamaus2020, Nadathur2019,Nadathur2020, Correa2022b}). 

However, void identification performed directly in redshift space introduces systematic effects that can strongly impact VSF and VGCF analyses. In particular, not only the voids shape is distorted by RSD and AP, but the LOS distortion can break the bijective correspondence between real- and redshift-space voids: some voids may be compressed below the finder resolution and consequently lost, while others may stretch and fragment into multiple void-in-void structures. Moreover, void centers are systematically displaced along the LOS, no longer tracing the real-space mass distribution \citep{Cai2016, Nadathur2019, Correa2022b}.
This motivates the need for void finders that remain robust in the presence of RSD and AP effects. 

Dynamical methods that exploit the full information of the mass field can alleviate these systematic effects without erasing the cosmological signal encoded in RSD and AP, providing an alternative to traditional reconstruction techniques that completely remove RSD information by restoring real-space positions \citep{Degni2025}. Nevertheless, recovering real-space tracer positions from reconstructed peculiar velocities remains crucial to simplify redshift-space analyses, isolate the AP contribution to the VGCF, and in general enable a more direct cosmological interpretation of void statistics. 

To validate our void finder under realistic survey conditions (light-cone geometry, realistic tracer sparseness, and effect of RSD) we apply it to the LRG sample from the Buzzard mock, presented in Sect. \ref{sec:buzzard}. First, we compare the void catalogs identified by \texttt{BitVF} and \texttt{REVOLVER} in both real and redshift space to assess whether \texttt{BitVF} intrinsically mitigates the detection systematic uncertainties introduced by RSD. Second, we study voids in the reconstructed real space obtained with \texttt{BitVF}, modeling RSD under the assumption of a fiducial cosmology (see Appendix~\ref{app:otderiv} for details on the OT-based correction).

\subsection{Redshift space}
\label{sec:zspace}

We apply \texttt{BitVF} and \texttt{REVOLVER} to the real- and redshift-space LRG samples of the selected Buzzard mock to assess their response to RSD-induced systematic effects in void identification.

A direct assessment of this behavior is provided by comparing the total number of voids and the uncleaned VSF in real and redshift space. The VGCF, instead, is not directly comparable between the two domains since RSD introduce anisotropies that are encoded in its quadrupole. However, it remains a useful cross-validation tool: consistency of the VGCF quadrupole between \texttt{BitVF} and \texttt{REVOLVER} ensures that the dynamical definition does not generate spurious anisotropies beyond those expected from RSD.

For this comparison, we configure \texttt{BitVF} with $N_{\mathrm{rec}}\! =\! 50$, $\epsilon_\mathrm{rec} \!=\! 10^{-3}$, $\sigma_{\mathrm{sm}} \!=\! 1\,\mathrm{MPS}$, and a grid spacing of $1/\sqrt[3]{2}\,\mathrm{MPS}$, consistent with our resolution prescriptions (Fig.~\ref{fig:resolution}) and designed to mimic the scale sensitivity of topological methods.
\begin{figure}
    \centering
    \includegraphics[width=1.0\linewidth]{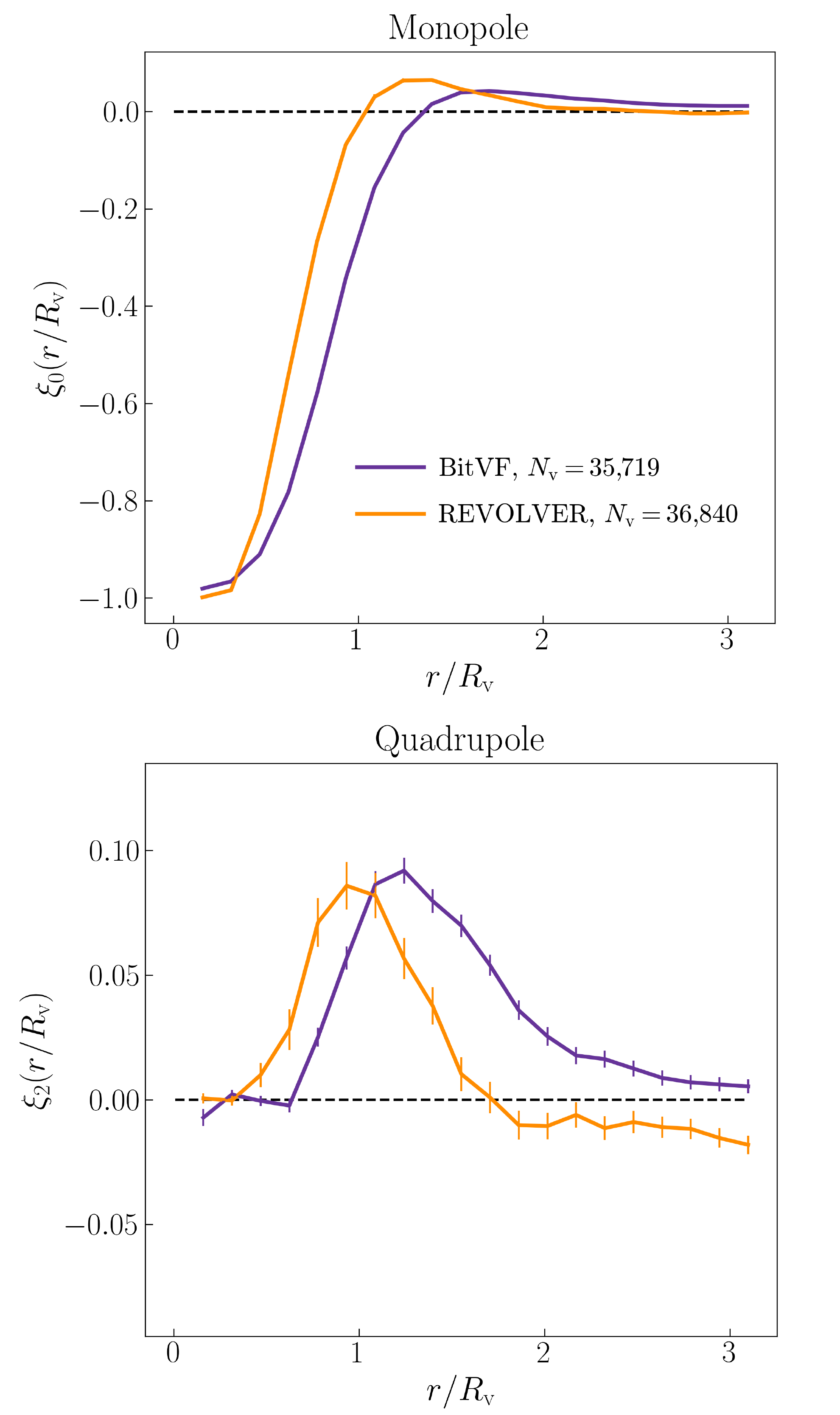}
    \caption{Monopole (\textit{top}) and quadrupole (\textit{bottom}) of the VGCF measured from the redshift-space catalogs of \texttt{BitVF} (purple) 
    and \texttt{REVOLVER} (orange) voids in the Buzzard mock. Radial separations are rescaled by the effective radius of each void before stacking. 
    The monopole exhibits the characteristic underdensity and compensation wall, with \texttt{BitVF} profiles appearing smoother and systematically 
    shifted toward larger rescaled radii compared to \texttt{REVOLVER}. The quadrupole shows the expected positive signal induced by 
    RSD for both void catalogs, with the same relative shift observed for the dynamical method. Errorbars are recovered from the diagonal of the covariance matrices.}
    \label{fig:ccZspace}
\end{figure}
\begin{figure*}
    \includegraphics[width=1.\linewidth]{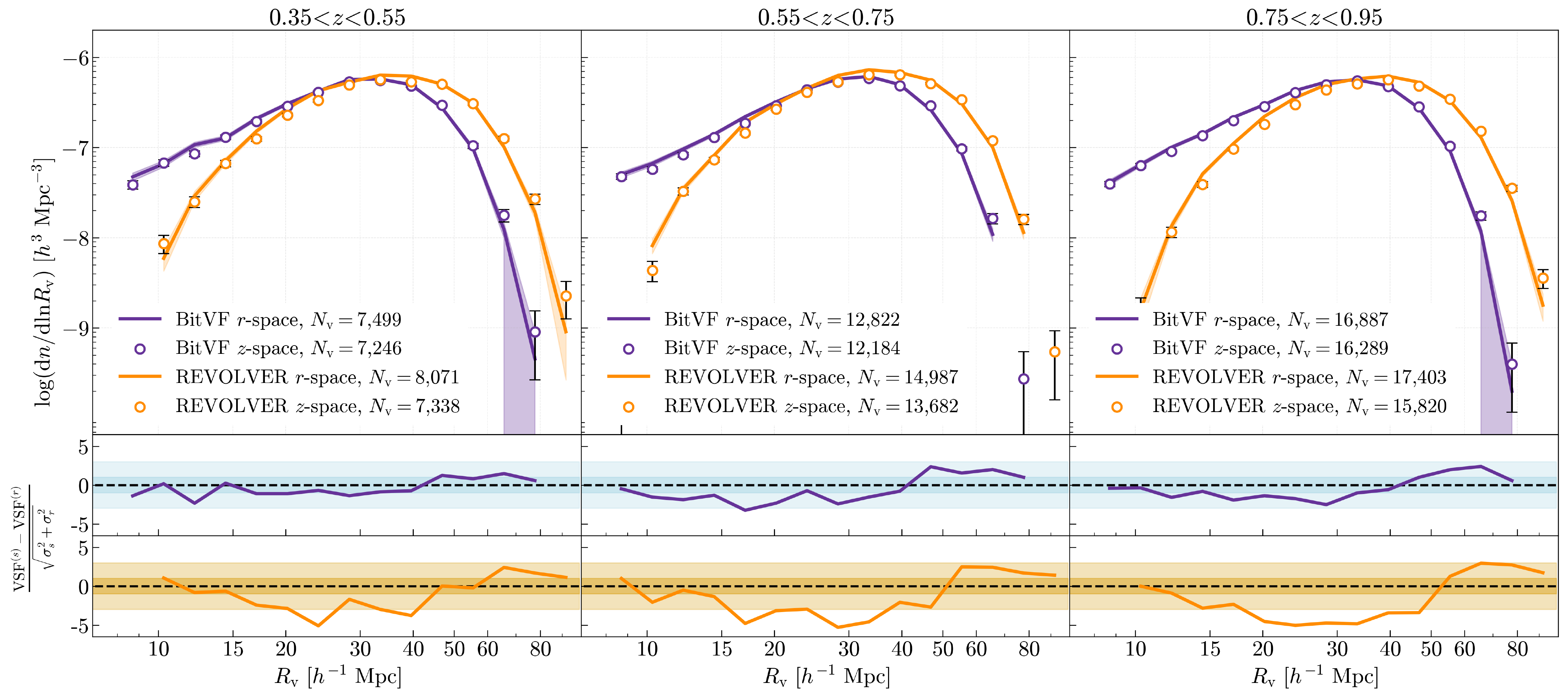}
    \caption{VSFs measured from real- and redshift-space catalogs in three equispaced redshift bins ($0.35\!<\!z\!<\!0.55$, $0.55\!<\!z\!<\!0.75$, $0.75\!<\!z\!<\!0.95$; 
    left, center, and right columns, respectively). The top panels show the VSFs obtained with \texttt{BitVF} (purple) 
    and \texttt{REVOLVER} (orange): solid lines indicate voids identified in real space, while dots report the corresponding redshift-space results. 
    Errorbars on the VSFs are Poissonian. The middle and bottom panels display the residuals with respect to the real-space catalog,
    with shaded bands marking statistical consistency levels: a darker band for the $1\sigma$ region and a lighter band for the $3\sigma$ region. The comparison highlights the 
    different response of the two methods to RSD. While \texttt{REVOLVER} exhibits significant discrepancies between real and redshift spaces, \texttt{BitVF} preserves 
    the void number and yields VSFs that remain consistent within $3\sigma$ across all redshift bins.
    }
    \label{fig:vsfRZ}
\end{figure*}
We begin by comparing the VGCF monopole and quadrupole of the full redshift-space catalogs (Fig.~\ref{fig:ccZspace}), computed using the Davis–Peebles estimator \citep{DavisPeebles1983}, for which we verified that differences with the Landy–Szalay estimator \citep{LandySzalay1993} are negligible within statistical uncertainties. The VGCF is measured as a function of the rescaled separation \(r/R_{\mathrm v}\), where \(R_{\mathrm v}\) is the effective radius of each void, in 20 linear bins over \(0 < r/R_{\mathrm v} < 3\). This rescaling allows voids of different sizes to be stacked and compared on a common relative scale, with uncertainties estimated from the diagonal of the covariance matrices.

The monopoles exhibit the characteristic underdensity and compensation wall, with \texttt{BitVF} profiles appearing smoother and shifted toward larger rescaled radii relative to \texttt{REVOLVER}, shift due to the different void definition between the methods. The quadrupoles show the anticipated positive signal induced by RSD for both catalogs, with a similar radial shift for \texttt{BitVF}, confirming that the two methods respond consistently to RSD.

We then quantify the RSD impact on void identification through the uncleaned VSF and void counts (Fig.~\ref{fig:vsfRZ}), computed in three equispaced redshift bins ($0.35\!<\!z\!<\!0.55$, $0.55\!<\!z\!<\!0.75$, $0.75\!<\!z\!<\!0.95$) to trace variations with redshift and bias. The number of voids in each bin is reported in the corresponding panel.
The comparison shows that \texttt{BitVF} substantially mitigates the effects of RSD on void statistics. The fraction of voids lost when moving from real to redshift space decreases to $3.2\%$, $5.0\%$, and $3.5\%$ in the three bins, compared to $9.1\%$, $8.7\%$, and $9.1\%$ for \texttt{REVOLVER}. Likewise, the VSFs of \texttt{BitVF} remain consistent within the $3\sigma$ level across all bins (and often within the $1\sigma$ level, especially at low-$z$), whereas the topological void finder exhibits discrepancies up to $5\sigma$.
These results demonstrate that the dynamical identification implemented in \texttt{BitVF} provides a more stable characterization of void statistics in redshift space, preserving void counts and VSF consistency despite the presence of RSD.

\subsection{Reconstructed space}
\label{sec:recspace}

A natural way to mitigate redshift-space related systematic effects is to reconstruct an approximate real-space distribution of tracers from their observed redshift-space positions by correcting for LOS peculiar velocities. In this work, these velocities are inferred from the displacement field recovered with the OT algorithm described in Sect.~\ref{reconstruction}.

To perform this correction, we modify the classic \citet{Kaiser1987} mapping between real- and redshift-space positions to include the linear bias present in the OT-reconstructed displacement field. This ensures that reconstructed real-space positions of tracers are consistently derived from their redshift-space coordinates. The full derivation is presented in Appendix~\ref{app:otderiv}.

\begin{figure}[h!]
    \centering
    \includegraphics[width=1.0\linewidth]{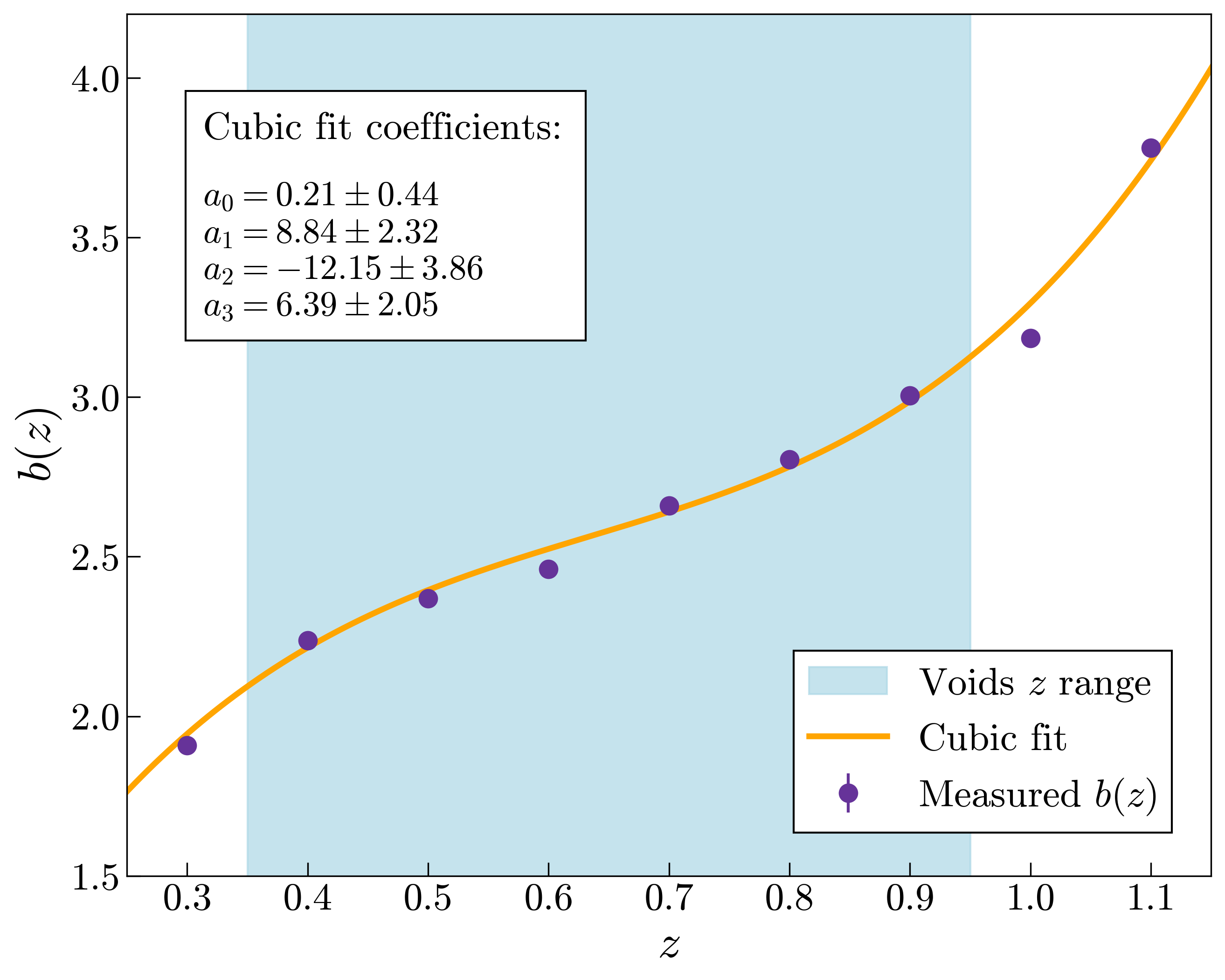}
    \caption{Linear bias measurements (purple dots) for the LRG population of the Buzzard mock in redshift bins of width \(\Delta z = 0.1\). 
    The cubic fit is shown in orange, with the fitted coefficients indicated. The light blue band marks the redshift range used for void identification.}
    \label{fig:bias}
\end{figure}

\begin{figure*}
    \includegraphics[width=1.\linewidth]{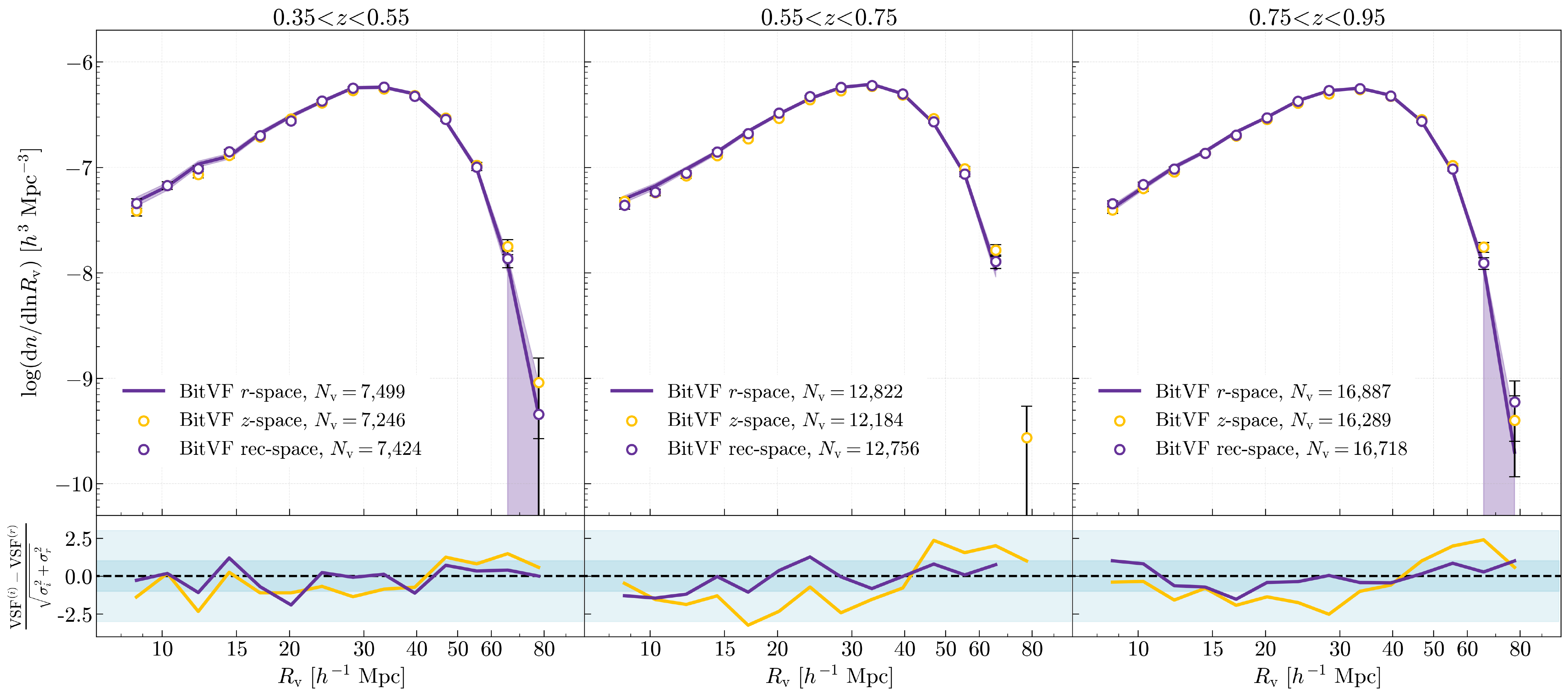}
    \caption{Comparison between VSFs in real (solid lines), redshift (saffron dots), and reconstructed (purple dots) spaces across three redshift bins. 
    Lower panels show residuals with respect to the real-space VSF, with shaded bands indicating the \(1\sigma\) and \(3\sigma\) confidence intervals. 
    The reconstructed-space VSF closely follows the real-space reference, demonstrating the effectiveness of the reconstruction.}
    \label{fig:vsfRecon}
\end{figure*}

In this framework, Eq.~\eqref{eq:velpsirel}, which relates the LOS velocity to the displacement field, becomes
\begin{equation}
\mathbf{v}_\parallel = a H \frac{f}{(b+f)} \mathbf{\Psi}_{\mathrm{OT}\parallel} ,
\end{equation}
where the factor \(1/(b+f)\) accounts for the linear bias introduced by the OT reconstruction (see Appendix \ref{app:otderiv}).  
Accordingly, the mapping from redshift to real space (Eq.~\ref{eq:originalxs}) reads
\begin{equation}
\label{eq:OTrspaceCorr}
\mathbf{r} = \mathbf{s} - \frac{f}{(b+f)} \mathbf{\Psi}_{\mathrm{OT}\parallel} \hat{l} .
\end{equation}
Since the OT reconstruction provides a back-in-time displacement field, the sign of \(\mathbf{\Psi}_{\mathrm{OT}}\) must be reversed when correcting redshift-space positions.

The reconstruction follows two steps. First, the OT algorithm provides an estimate of the tracer displacement field, which is then used to correct their positions through Eq.~\eqref{eq:OTrspaceCorr}. Second, the void finder is run on the reconstructed catalog using the same configuration adopted for the real- and redshift-space analyses: \(N_{\mathrm{rec}} = 50\), \(\epsilon_\mathrm{rec} = 10^{-3}\), \(\sigma_{\mathrm{sm}} = 1\,\mathrm{MPS}\), and \(l_\mathrm{cell}=1/\sqrt[3]{2}\,\mathrm{MPS}\).

The linear growth rate $f$ is computed assuming the Buzzard cosmology (Sect.~\ref{sec:buzzard}), 
while the linear bias $b$ is measured in nine redshift bins between $z\! =\! 0.25$ and $z \!=\! 1.15$ 
using the standard real-space two-point correlation function (2PCF) method \citep{Kaiser1984}. We computed the bias value in each bin following the method described in Appendix \ref{app:velcomp}.
A cubic polynomial is then fitted to the \(b(z)\) measurements, as shown in Fig.~\ref{fig:bias}.

Once the reconstruction is performed, we identify voids in the reconstructed real space using our dynamical void finder.
To test the accuracy of the reconstruction, we compare these voids with those identified directly in the real space of the mock.
Specifically, we examine whether the VSF of the reconstructed voids statistically matches its real-space counterpart, and whether the quadrupole of the VGCF is effectively suppressed, as expected for a correct recovery of real-space positions.

Figure~\ref{fig:vsfRecon} presents the comparison between real- and reconstructed-space VSFs. Across all redshift bins, the agreement is excellent: most measurements lie within the \(1\sigma\) uncertainty, and none deviate beyond \(2\sigma\). The total void counts further confirm this consistency, the reconstructed catalog yields void numbers within \(1\%\) of the real-space reference, a substantial improvement over redshift space, where RSDs distort void shapes, shift the VSF toward larger radii, and suppress small voids. Thus, the OT-based reconstruction nearly restores the real-space void abundance.

\begin{figure*}
    \centering
    \includegraphics[width=1.\linewidth]{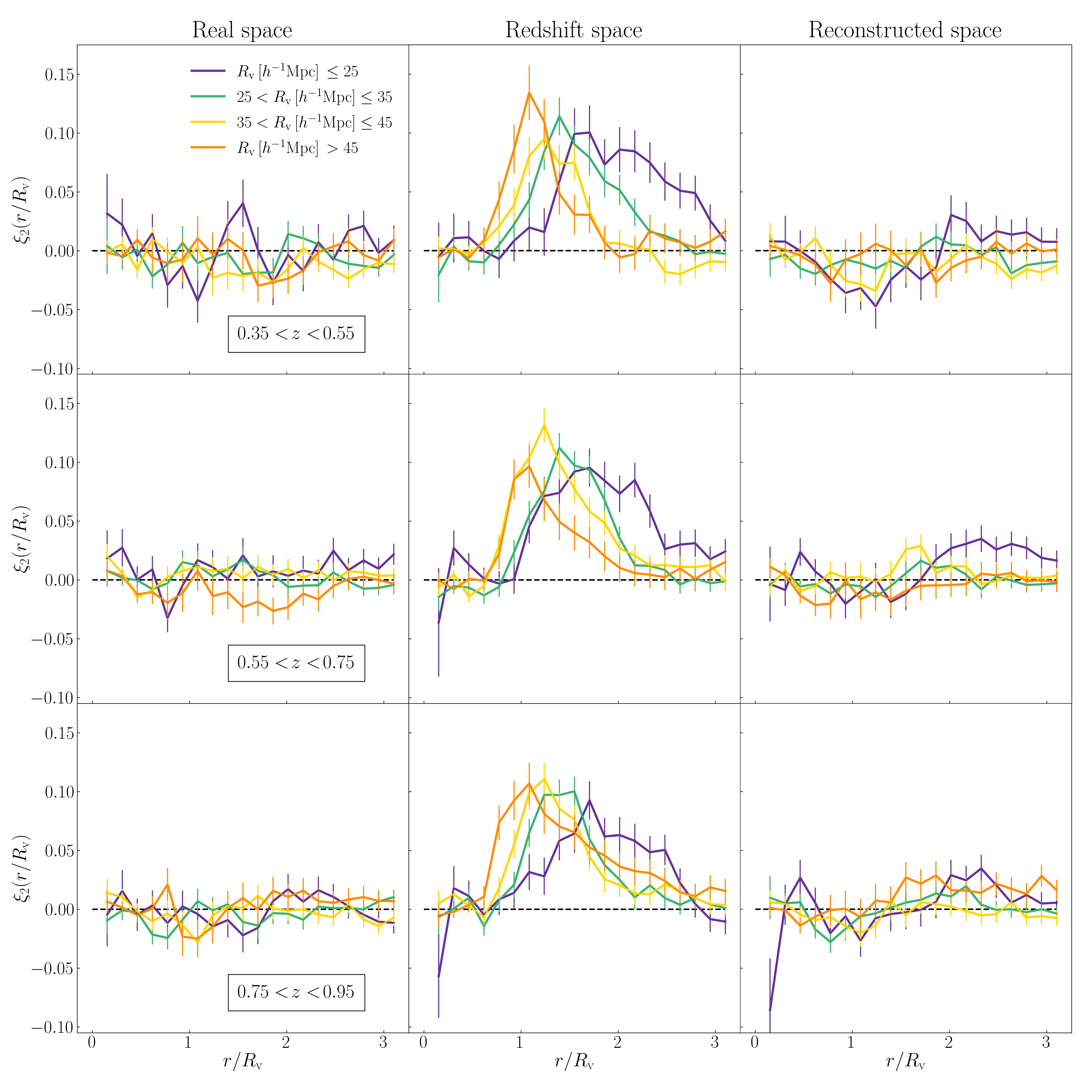}
    \caption{Quadrupoles of the VGCF in three redshift bins (rows) for real (left), redshift (middle), and reconstructed (right) spaces. 
    Real-space quadrupoles show the expected isotropy, while redshift-space ones exhibit the anisotropies induced by peculiar velocities. 
    Reconstructed-space quadrupoles are consistent with zero, confirming the effective correction of RSDs, with only minor residuals for the smallest voids.
    Errorbars are taken from
the diagonal of the associated covariance matrices.}
    \label{fig:recQuad}
\end{figure*}

A complementary test is provided by the quadrupole of the VGCF, shown in Fig.~\ref{fig:recQuad}. The reconstructed-space quadrupoles are consistent with zero across all redshift bins, matching the real-space results, indicating that our OT reconstruction well restored the isotropy, mitigating RSD effects. Small residuals appear for the smallest voids, likely due to statistical noise, edge effects, or the limited resolution of the OT displacement field. Overall, the analysis confirms that the OT-based correction effectively removes the anisotropies induced by RSD.

In summary, our reconstruction method, based on the OT displacement field and a bias-corrected \citet{Kaiser1987} mapping, successfully recovers the main statistical properties of real-space voids. Both the VSF and the VGCF quadrupole are restored to their real-space forms, with only sub-percent residuals at small scales. This demonstrates that the OT-based reconstruction efficiently mitigates RSD and yields a reliable real-space void catalog across all redshifts, suitable for cosmological applications.

\section{Conclusions}
\label{sec:conclusions}

In this work, we have presented and tested a novel dynamical void finder, \texttt{BitVF}, 
which identifies voids as non-overlapping regions where the divergence of the back-in-time Lagrangian displacement field, reconstructed through 
an OT approach, is negative. This dynamical criterion provides a physically motivated characterization of voids as expanding 
basins in the primordial matter distribution, naturally connected to large-scale gravitational dynamics. Moreover, a dynamical void identification is expected to largely 
mitigate shot-noise issues that affect standard density-based and topological methods, which rely on directly sampling the density field in 
underdense regions, where mass tracers are sparse.

The algorithm proceeds in three steps: 
(i) OT-based reconstruction of the displacement field from present-day tracers back to their Lagrangian positions;
(ii) computation of the divergence on a regular grid; (iii) segmentation into distinct basins of the regions with negative divergence of the reconstructed displacement field through a watershed algorithm.
Technical aspects such as interpolation, smoothing, and boundary treatment were carefully addressed to ensure robust performances. We validate the OT reconstruction by showing good agreement between reconstructed and true halo velocities in the Aletheia simulation (Appendix~\ref{app:velcomp}).
Systematic tests, presented in Appendix \ref{app:impact} demonstrated the stability of \texttt{BitVF} across a broad range of reconstruction parameters, including the number of iterations, 
grid resolution, smoothing scale, and convergence threshold.

In Sect. \ref{sec:tests} we then validated the method against a standard topological void finder, namely \texttt{REVOLVER} \citep{Revolver2019}, using halo catalogs from the Aletheia simulation box. 
In this comparison, we observed consistent VSFs once the different void definitions were properly taken into account: \texttt{BitVF} naturally 
identifies smaller structures due to its dynamical criterion, which selects only regions characterized by mass outflow, while \texttt{REVOLVER} fills the 
entire volume with voids. This difference leads to a rescaled VSF relative to the topological case. At the same time, the density profiles 
of \texttt{BitVF} voids appeared smoother and more regular, providing a clear indication of a coherent identification that accurately reflects the underlying 
large-scale structure evolution.

Moreover, we performed a stability tests on different subsample of the Aletheia  dark matter field, evaluating the behavior of the algorithm under limiting conditions such as aggressive 
subsampling and high density of the sample, which can degrade the precision of the void identification process. These tests confirmed that the dynamical criterion adopted by \texttt{BitVF} is intrinsically less sensitive to shot noise than traditional 
density-based approaches, with cleaned VSF remaining stable across various degrees of sparseness. This demonstrates the advantage 
of a void finding definition which relies on reconstructed dynamics rather than on the direct sampling of underdense regions.

Finally, in Sect. \ref{sec:lightcone} we investigated the impact of RSD using a realization of the LRGs DESI-like Buzzard mock catalog. Applying \texttt{BitVF} in 
redshift space reduced the differences in the VSF relative to real space and preserved the total number of voids more effectively 
than topological methods. Using the OT-reconstructed displacement field with a corrected \citet{Kaiser1987} mapping, presented in Appendix \ref{app:otderiv}, we recovered 
reconstructed-space void catalogs in which both the void abundance and the quadrupole of the void-galaxy correlation function are consistent 
with the real-space reference, demonstrating a robust mitigation of RSD-induced systematics effects.

Overall, \texttt{BitVF} provides a physically motivated, dynamical alternative to traditional topological void-finding algorithms, establishing a clear 
connection between the present-day large-scale structure and the initial Lagrangian conditions. This dynamical approach mitigates RSD-induced systematic effects on void identification in redshift space and proved to effectively recover the real-space void statistics exploiting the underlying OT reconstruction, once the true cosmology is assumed. While practical choices such as smoothing, random density 
sampling, and small-scale reconstruction affect the identification of the smallest voids, these effects are well understood and do not compromise 
the overall performance. The \texttt{BitVF} and OT reconstruction codes will be publicly released as part of a dedicated \texttt{GitLab} package 
and included in the \texttt{CosmoBolognaLib} \citep{cbl}, enabling reproducible applications in cosmology.
\begin{acknowledgements}
SS would like to thank Enzo Branchini and Ravi Sheth for insightful discussions on the theoretical framework, development, and future perspectives of this work. SS would also like to thank Chris Blake and Joe DeRose for their efforts in the creation of the Buzzard galaxy mocks and for their generous availability, as well as Matteo Esposito for the production of the Aletheia simulations and for kindly providing access to them.

GD acknowledges support from the french government under the France 2030 investment plan, as part of the Initiative d’Excellence d’Aix- Marseille Université - A*MIDEX AMX-22-CEI- 03.

FM and LM acknowledge the financial contribution from the 
PRIN-MUR 2022 20227RNLY3 grant “The concordance cosmological model: stress-tests with galaxy clusters” supported by Next Generation EU and from the grant ASI n. 2024-10-HH.0 “Attività scientifiche per la missione Euclid – fase E”.

This work was performed using the Dark Energy Center (DEC) hosted at Aix Marseille Univ, CNRS/IN2P3, CPPM, Marseille, France.
\end{acknowledgements}

\bibliography{bibliography}
\appendix
\section{Validation of the reconstructed velocity field}
\label{app:velcomp}
To validate the accuracy of the reconstructed displacement field, we applied the reconstruction algorithm to the Aletheia halo catalog described in Sect.~\ref{sec:aletheia}. 
The test was performed using 50 random realizations and imposing a convergence threshold for the minimization of the cost function, $\epsilon_\mathrm{rec}\! =\! 10^{-3}$. 
Convergence is defined as the condition in which, over a full iteration of the algorithm, fewer than one pair swap occurs per $10^{3}$ pairs considered.

Reconstructed velocities are obtained from the reconstructed displacement field through Eq.~\eqref{eq:velpsirel} and corrected for the linear bias. 
They are then compared with the true halo velocities extracted directly from the simulation.

The linear bias $b$ is estimated as
\begin{equation}
b = \sqrt{\frac{\xi_\mathrm{t}(r)}{\xi_\mathrm{m}(r)}} \quad \text{for} \quad r \gtrsim 20\,h^{-1}\mathrm{Mpc} \, ,
\end{equation}
where the tracer two-point correlation function $\xi_\mathrm{t}(r)$ is measured using the Landy--Szalay estimator \citep{LandySzalay1993} as implemented in the \texttt{CosmoBolognaLib} package \citep{cbl}, while the dark matter two-point correlation function $\xi_\mathrm{m}(r)$ is derived from the linear matter power spectrum computed with \texttt{CAMB} \citep{Lewis2000}. 
Large scales are used to ensure linearity \citep{Kaiser1984}.

\begin{figure}[h]
    \centering
    \includegraphics[width=1.\linewidth]{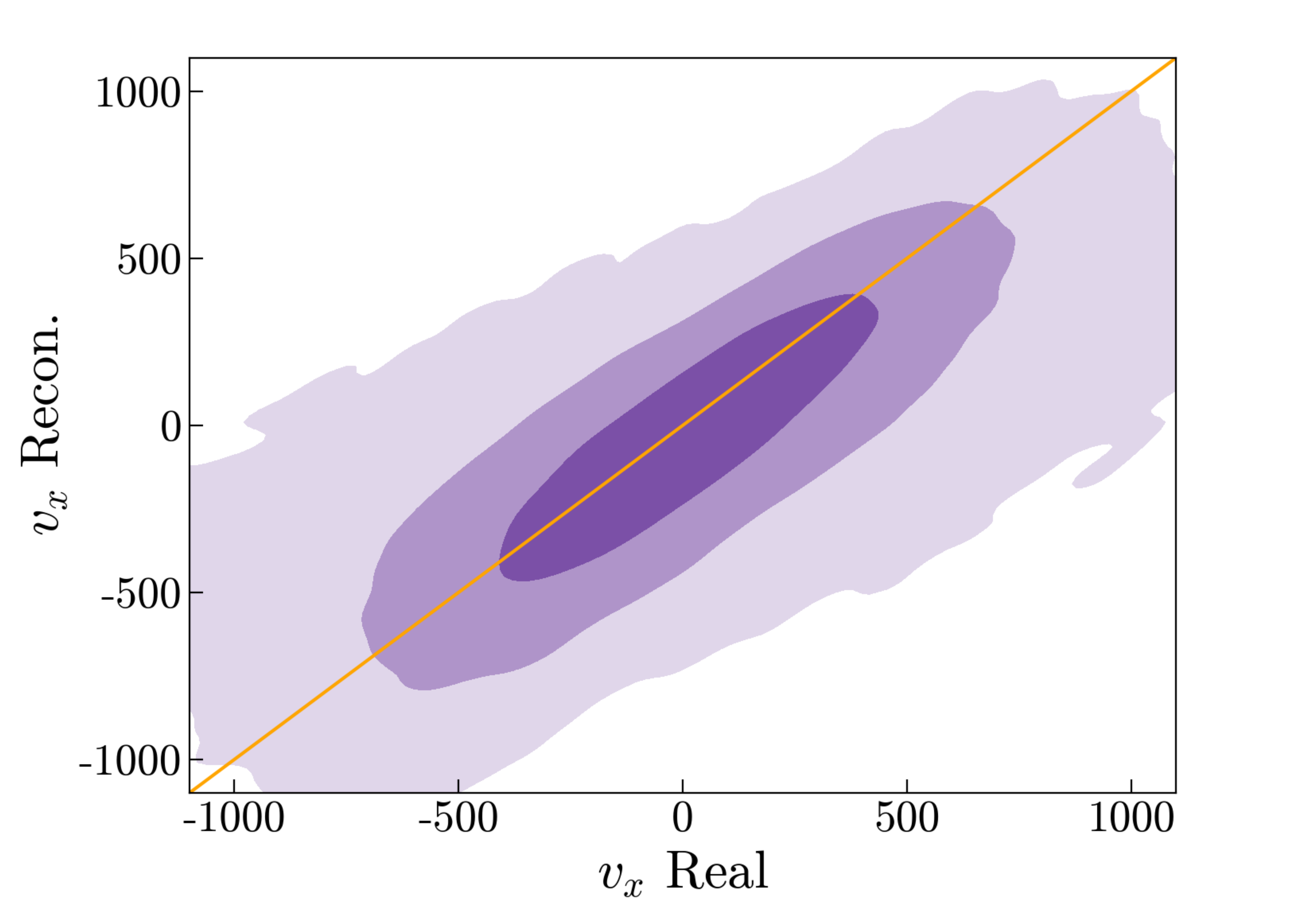}
    \caption{Contours in the reconstructed versus true velocity space for the $x$-component of halo velocities in the Aletheia simulation. Contours correspond to 68.27\%, 95.45\% and 99.73\% of the distribution. The diagonal line indicates the ideal one-to-one correspondence between reconstructed and true velocities.}
    \label{fig:vCon}
\end{figure}

Figure~\ref{fig:vCon} shows the comparison between reconstructed and true halo velocities along the $x$-axis; the remaining Cartesian components show qualitatively similar behavior. 
We find a Pearson correlation coefficient of $R \simeq 0.77$, indicating good agreement between the reconstructed and true velocity fields. 
Residual discrepancies are primarily due to physical limitations of the reconstruction, such as small-scale stochastic motions and shell-crossing effects, which cannot be captured by an OT-based approach.

\section{Evaluation of the impact of the internal parameters}
\label{app:impact}

The performance of \texttt{BitVF} depends on a set of internal parameters controlling both the OT reconstruction of the displacement field and the computation of its divergence, which together determine the void identification. While the conceptual framework of the algorithm is fixed, the resulting void statistics, such as their abundance and density profiles, can vary with these parameter choices. Assessing their influence is therefore crucial to establish the method robustness and to guide its application to real data.

In this Appendix, we focus on four parameters: \(N_{\mathrm{rec}}\), \(\epsilon_\mathrm{rec}\), $l_\mathrm{cell}$ and \(\sigma_\mathrm{sm}\). These naturally form two parameter pairs: \((N_{\mathrm{rec}}, \epsilon_\mathrm{rec})\), controlling the accuracy and convergence of the reconstruction, and ($l_\mathrm{cell}$, \(\sigma_\mathrm{sm}\)), defining the effective resolution of the divergence field.

We explore this parameter space by varying one pair at a time while keeping the other fixed, and we evaluate the resulting variations in the VSF  and in the stacked density profiles. This procedure isolates the sensitivity of each parameter set and identifies stable regions where results remain consistent.
In addition, we discuss the choice of grid resolution. The minimum cell size is constrained not only by computational cost but also by the requirement that each cell contains a sufficient number of random tracers, typically 20–30 on average, to ensure that shot-noise fluctuations are well approximated by Gaussian statistics. 
All tests are performed using the halo sample from the Aletheia simulation (Sect.~\ref{sec:aletheia}). A border of \(30\,h^{-1}\mathrm{Mpc}\) is removed from the box boundaries to minimize the impact of truncated voids.

\subsection{Reconstruction sensitivity}

We first explore the pair \((N_\mathrm{rec}, \epsilon_\mathrm{rec})\), which jointly determines reconstruction accuracy and convergence of the OT solution. We scan the two-dimensional parameter space spanned by \(N_\mathrm{rec}\) and \(\epsilon_\mathrm{rec}\) and evaluate the resulting void catalogs through two observables: the uncleaned VSF and the stacked VGCF monopole.

To isolate the effect of each parameter, we consider two one-dimensional cuts: (i) fixing $\epsilon_\mathrm{rec}=0$ and varying $N_\mathrm{rec}$ between 1 and 50, and (ii) fixing $N_\mathrm{rec}=50$ and varying $\epsilon_\mathrm{rec}$ between 1 (no swapping) and 0 (requirement of maximum precision). The baseline configuration is the most accurate one, $N_\mathrm{rec}=50$ and $\epsilon_\mathrm{rec}=0$.  
All tests are performed without Gaussian smoothing and using a grid-cell size of $l_\mathrm{cell}=1\,\mathrm{MPS}=7.81\,h^{-1}\mathrm{Mpc}$.
By retaining small-scale modes, which are the most challenging to reconstruct, we deliberately push the algorithm toward its operational limits. This choice allows potential instabilities or sensitivities to parameter choices to manifest more clearly.

\begin{figure*}
    \centering    
    \includegraphics[width=1\linewidth]{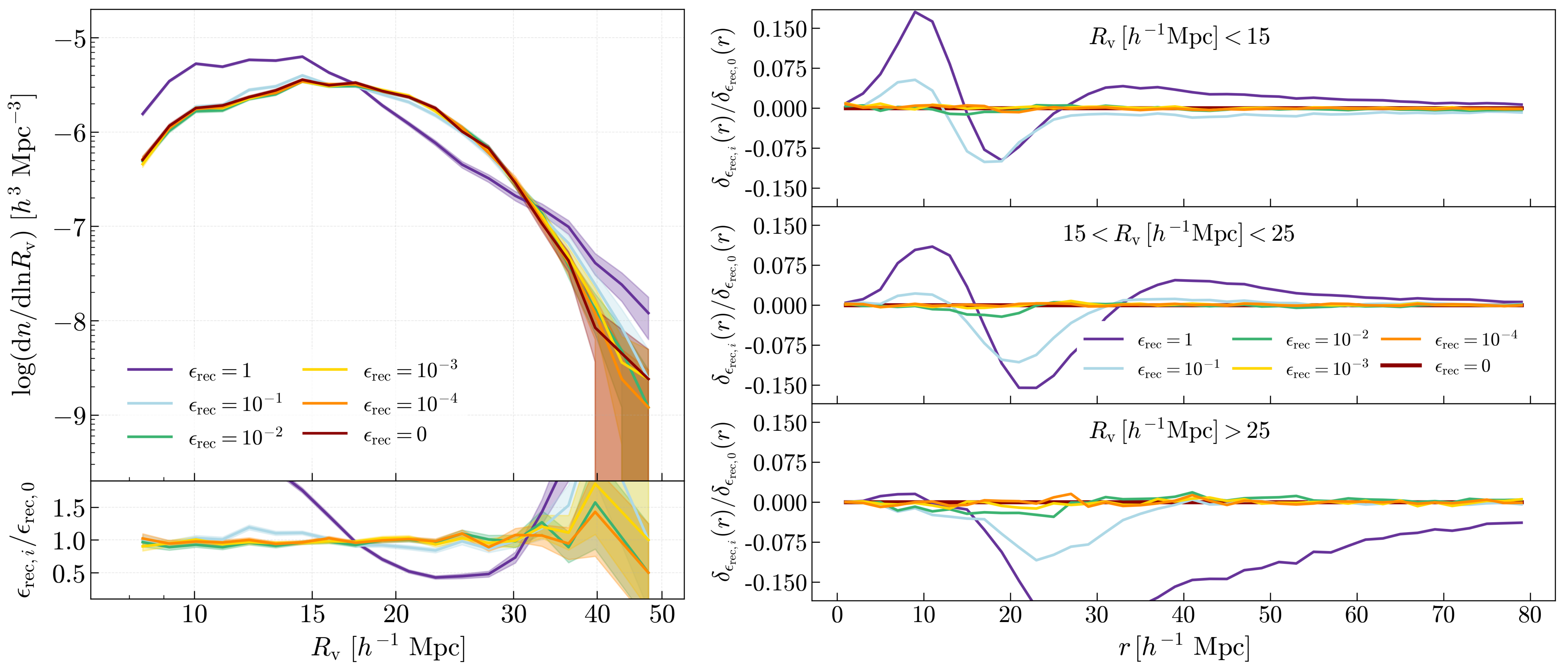}
    \caption{Impact of the convergence threshold variation between 1 and 0 on the void catalogs, with fixed $N_\mathrm{rec}=50$.
    \textit{Left}: VSFs shown as solid lines in the upper subpanel, with relative Poisson uncertainties indicated by colored bands. The lower subpanel shows the ratios with respect to the reference configuration ($\epsilon_\mathrm{rec}=0$, $N_\mathrm{rec}=50$), with uncertainties obtained by propagation of the VSF errors.
    \textit{Right}: Ratios of the VGCF monopole with respect to the reference configuration, computed for different void radius bins to isolate the impact of the threshold on small, intermediate, and large voids.
    }
    \label{fig:thTest}
\end{figure*}

\begin{table}[h]
    \captionof{table}{Void counts for different $N_\mathrm{rec}-\epsilon_\mathrm{rec}$ combinations.}
    \centering
    \resizebox{0.5\textwidth}{!}{
    \begin{tabular}{c|cccccc}
    \specialrule{.15em}{.2em}{0.2em} 
    \specialrule{.05em}{.05em}{0.5em} 
    \textbf{$\mathbf{N_\mathrm{rec}}\backslash\epsilon_\mathrm{rec}$} & 1 & $10^{-1}$ & $10^{-2}$ & $10^{-3}$ & $10^{-4}$ & 0 \\
    \specialrule{.05em}{.5em}{0.5em} 
    1 & 71,452 & 65,364 & 58,782 & 57,717 & 57,211 & 57,049 \\[0.5em]
    2 & 71,698 & 63,164 & 54,910 & 53,722 & 53,464 & 53,381 \\[0.5em]
    5 & 69,886 & 57,357 & 47,911 & 46,990 & 46,830 & 46,602 \\[0.5em]
    10 & 65,976 & 48,774 & 40,854 & 40,613 & 40,270 & 40,249 \\[0.5em]
    20 & 60,814 & 40,803 & 35,560 & 35,406 & 35,683 & 35,704 \\[0.5em]
    30 & 57,082 & 36,827 & 33,324 & 33,413 & 33,891 & 34,148 \\[0.5em]
    50 & 51,647 & 33,098 & 31,308 & 31,841 & 32,093 & 32,499 \\
    \specialrule{.1em}{0.5em}{0.8em} 
    \end{tabular}
    }
    \label{tab:voidsThNrec}
\end{table}

Table~\ref{tab:voidsThNrec} summarizes the total void counts obtained for all tested parameter combinations. At fixed \(N_\mathrm{rec}\), the number of identified voids systematically decreases as the reconstruction threshold \(\epsilon_\mathrm{rec}\) is reduced, eventually reaching a stable plateau for \(\epsilon_\mathrm{rec} \lesssim 10^{-2}\text{-}10^{-3}\). This trend indicates that the reconstruction becomes effectively converged in this regime. Conversely, for fixed \(\epsilon_\mathrm{rec}\), increasing \(N_\mathrm{rec}\) also reduces the total void count, with a stronger suppression at higher thresholds. This behavior reflects the interplay between the two parameters: a larger number of random catalogs and a tighter convergence criterion both enhance the reconstruction accuracy, thereby mitigating the impact of shot noise and removing spurious small-scale voids.

The computational cost is largely dominated by the particle-swapping phase of the reconstruction, whose runtime increases rapidly as \(\epsilon_\mathrm{rec}\!\rightarrow\!0\). Establishing the minimum threshold ensuring convergence at a moderate computational expense is thus crucial to achieve a practical balance between precision and efficiency.

Figure~\ref{fig:thTest} shows that the VSF converges rapidly for $\epsilon_\mathrm{rec}\lesssim10^{-2}$, with only minor variations at smaller thresholds. The swapping step has a major influence on the results, particularly between $\epsilon_\mathrm{rec}=1$ and $\epsilon_\mathrm{rec}=10^{-1}$, where the reconstruction remains far from converging. Stacked VGCFs, computed by dividing the void catalogs into three size bins (small, intermediate, and large) to explore the dependence of the average density profiles on void size, exhibit stable profiles for $\epsilon_\mathrm{rec}\leq10^{-3}$. Larger thresholds yield void cores that appear artificially dense and compensation walls that are less pronounced, highlighting residual inaccuracies in the reconstruction. Moderate $\epsilon_\mathrm{rec}$ values therefore provide an optimal compromise between precision and computational efficiency.

\begin{figure*}[h]
    \centering    
    \includegraphics[width=1.\linewidth]{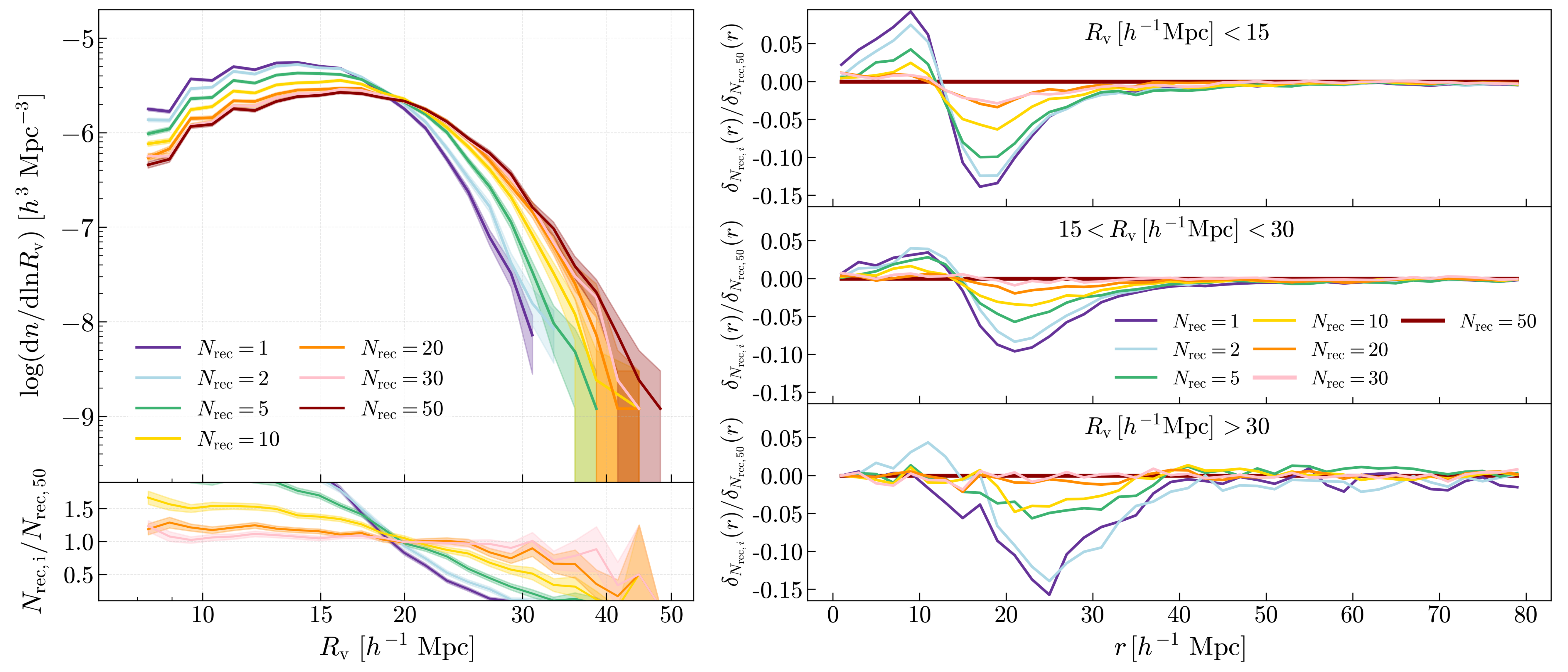}
    \caption{Impact of different numbers of random reconstructions ($N_\mathrm{rec}=1$-50) on the void catalogs, with fixed $\epsilon_\mathrm{rec}=0$.
\textit{Left}: VSFs shown in the upper subpanel, with relative Poisson uncertainties indicated by colored bands, and ratios with respect to the reference configuration in the lower subpanel, with uncertainties obtained by propagation of the VSF errors.
\textit{Right}: Ratios of the VGCF monopole with respect to the reference configuration, shown for different void radius bins.}
    \label{fig:NRECTest}
\end{figure*}

When $\epsilon_\mathrm{rec}=0$ and $N_\mathrm{rec}$ is varied (Fig.~\ref{fig:NRECTest}), convergence is slower. The VSF and VGCFs improve steadily up to $N_\mathrm{rec}=50$, especially for small voids. Large voids converge near $N_\mathrm{rec}\sim20$, intermediate ones near $\sim30$. Increasing $N_\mathrm{rec}$ suppresses shot-noise fragmentation, shifting the VSF toward larger void sizes and producing deeper, more compensated profiles.

\begin{figure}[h]
    \centering
    \includegraphics[width=1.\linewidth]{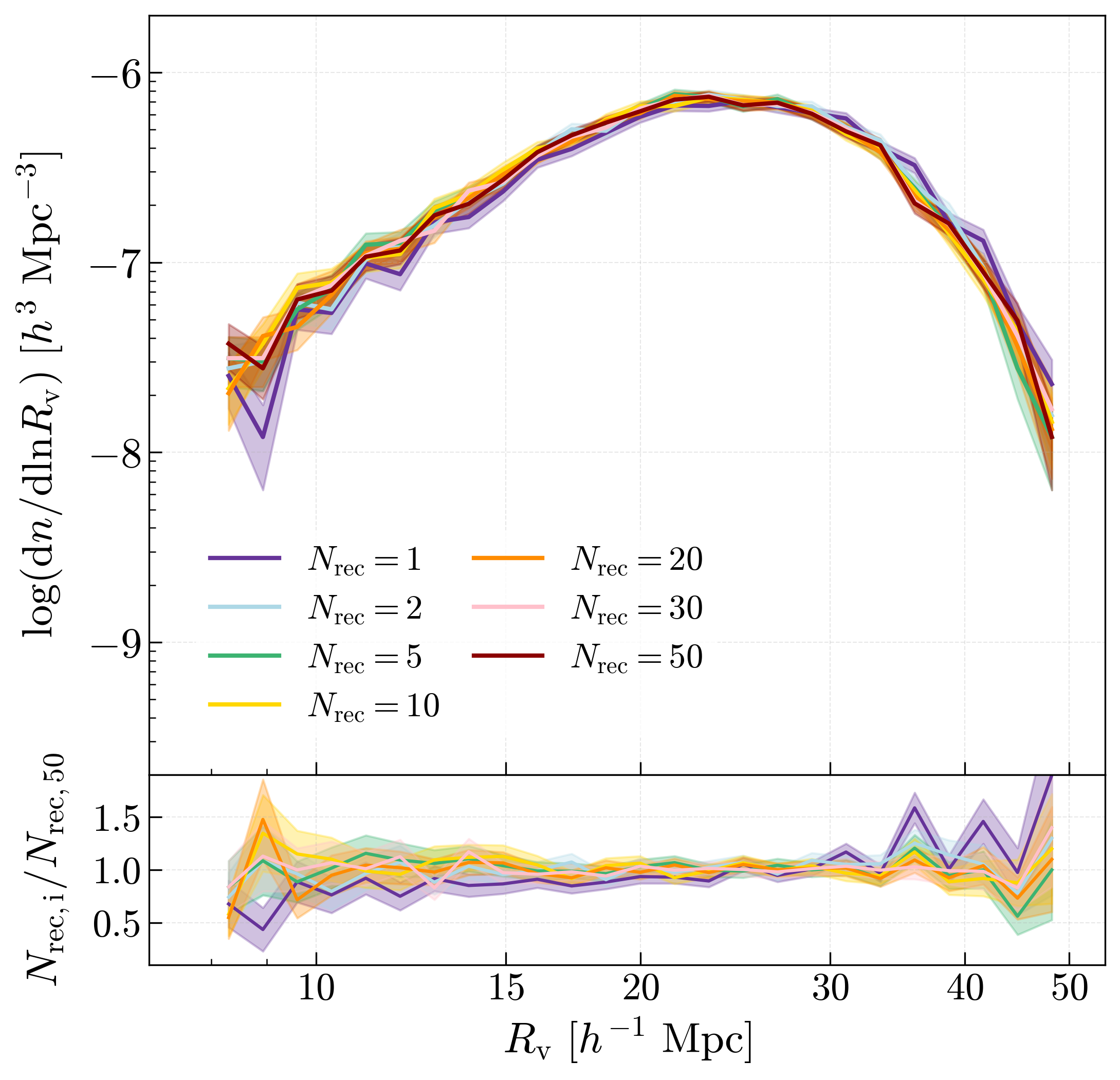}
    \caption{Impact of different numbers of random reconstructions ($N_\mathrm{rec}=1$-50) on the void catalogs, with fixed $\epsilon_\mathrm{rec}=0$, when a Gaussian smoothing with $\sigma_\mathrm{sm}=1\,\mathrm{MPS}$ is applied to the divergence field.
The upper subpanel shows the VSFs, with relative Poisson uncertainties indicated by colored bands, while the lower subpanel displays the ratios with respect to the reference configuration, with uncertainties obtained by propagation of the VSF errors.
Convergence is achieved with fewer random realizations compared to the case without Gaussian smoothing.
}
    \label{fig:sm1test}
\end{figure}

However, applying a Gaussian smoothing of $\sigma=1\,\mathrm{MPS}$ (Fig.~\ref{fig:sm1test}) yields rapid convergence with only a few random realizations, as small-scale modes, most affected by noise, are filtered out. This behavior is expected: without filtering, small-scale modes dominate and require a very dense random sampling to be reconstructed reliably. The same requirement naturally extends to survey-like situations, where the presence of masks and complex geometries further complicates the sampling.

Under the idealized conditions of our tests, convergence of both void counts and density profiles is reached with approximately ten random realizations and a stopping threshold of $10^{-2}$ when a Gaussian smoothing of $1\,\mathrm{MPS}$ is applied. In the absence of smoothing, however, the statistical properties of small voids remain sensitive to the density of the random catalog, requiring up (or more) to $N_\mathrm{rec}\!\simeq\!50$ to achieve stable results. The stopping threshold mainly influences the reconstruction at large values, while both the VSF and VGCFs stabilize for $\epsilon_\mathrm{rec}\!\lesssim\!10^{-2}$. In realistic survey applications, where shot noise, selection effects, and complex geometries can degrade the reconstruction, denser random catalogs and stricter convergence criteria may therefore be necessary to ensure robustness.

\subsection{Resolution of the Divergence Field}

The resolution of the divergence field is set by the smoothing length $\sigma_\mathrm{sm}$ and by the grid-cell size, $l_\mathrm{cell}$, which acts as complementary scale filter. $l_\mathrm{cell}$ fixes the intrinsic resolution of the discrete field, while $\sigma_\mathrm{sm}$ regulates the physical scale at which fluctuations are suppressed. Their combined effect directly shapes the number, size, and stability of the identified voids.

A theoretical lower limit on the resolution follows from sampling arguments: for a given tracer density, voids smaller than twice the MPS cannot be robustly identified \citep{Shannon1948,Ronconi2017}. However, this criterion is non trivial to apply directly, as it requires local sampling information.  

\begin{figure}[h]
    \centering
    \includegraphics[width=1\linewidth]{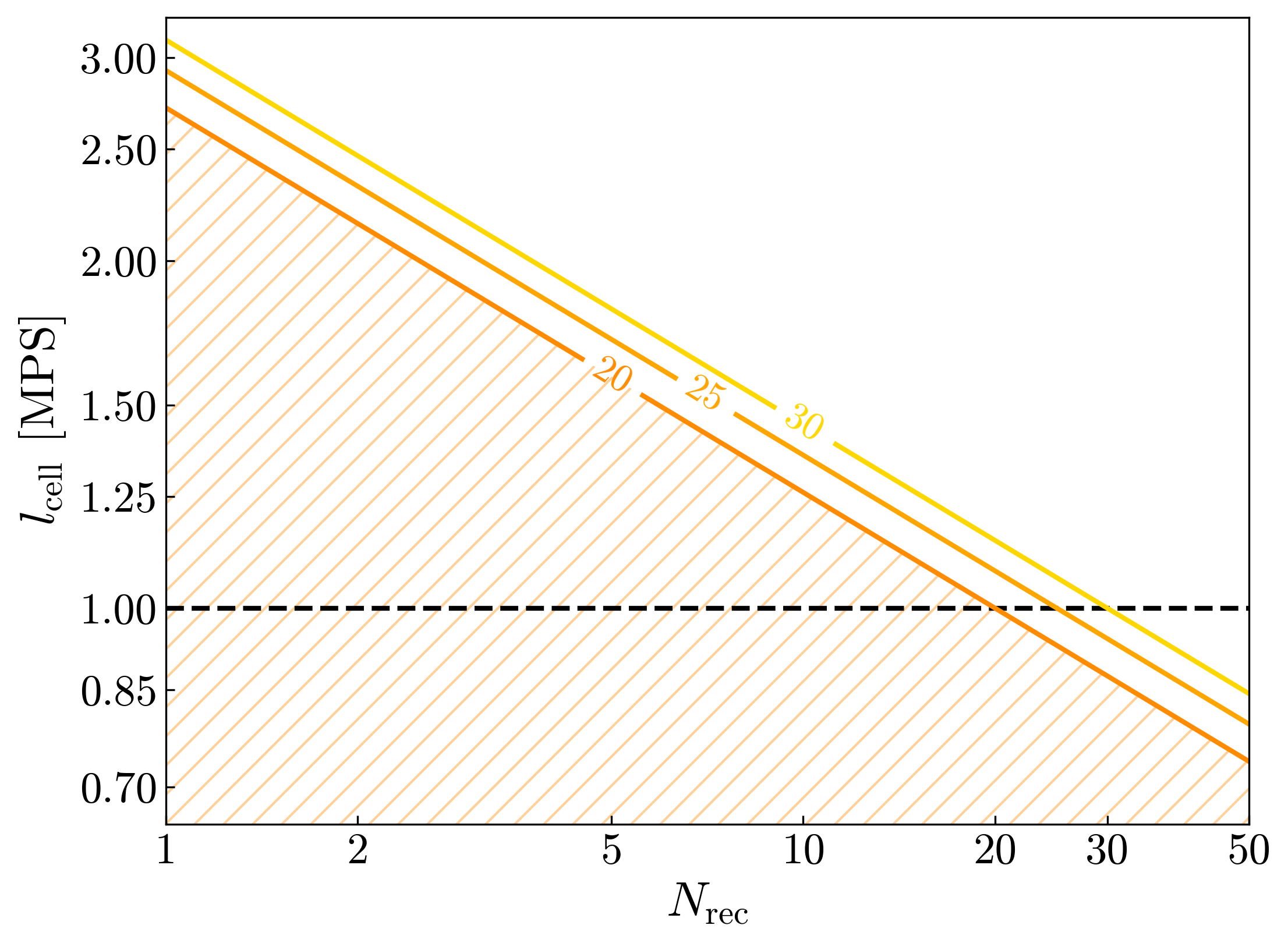}
    \caption{Grid cell size $l_\mathrm{rec}$, in units of MPS, as a function of the number of independent reconstructions $N_\mathrm{rec}$, each obtained using a different random realization and therefore setting the effective ratio between random particles and tracers. The figure shows the theoretical minimum resolution required for the cell occupancies to be well approximated by a Gaussian distribution. The orange dashed region marks the forbidden parameter space, while the reference lines correspond to and average of 20, 25, and 30 random particles per cell.}
    \label{fig:resolution}
\end{figure}

A more practical prescription relies on the density of the random catalogs: to ensure Gaussian shot-noise statistics, each grid cell should contain at least $\sim20$–30 random tracers, on average. Figure~\ref{fig:resolution} illustrates the resulting limits on cell size for different random densities.

We then test the impact of $\sigma_\mathrm{sm}$ and $l_\mathrm{cell}$ on the uncleaned VSF, adopting $N_\mathrm{rec}=50$ and $\epsilon_\mathrm{rec}=10^{-3}$ to minimize reconstruction-induced noise.

\begin{figure}
    \centering
    \includegraphics[width=0.98\linewidth]{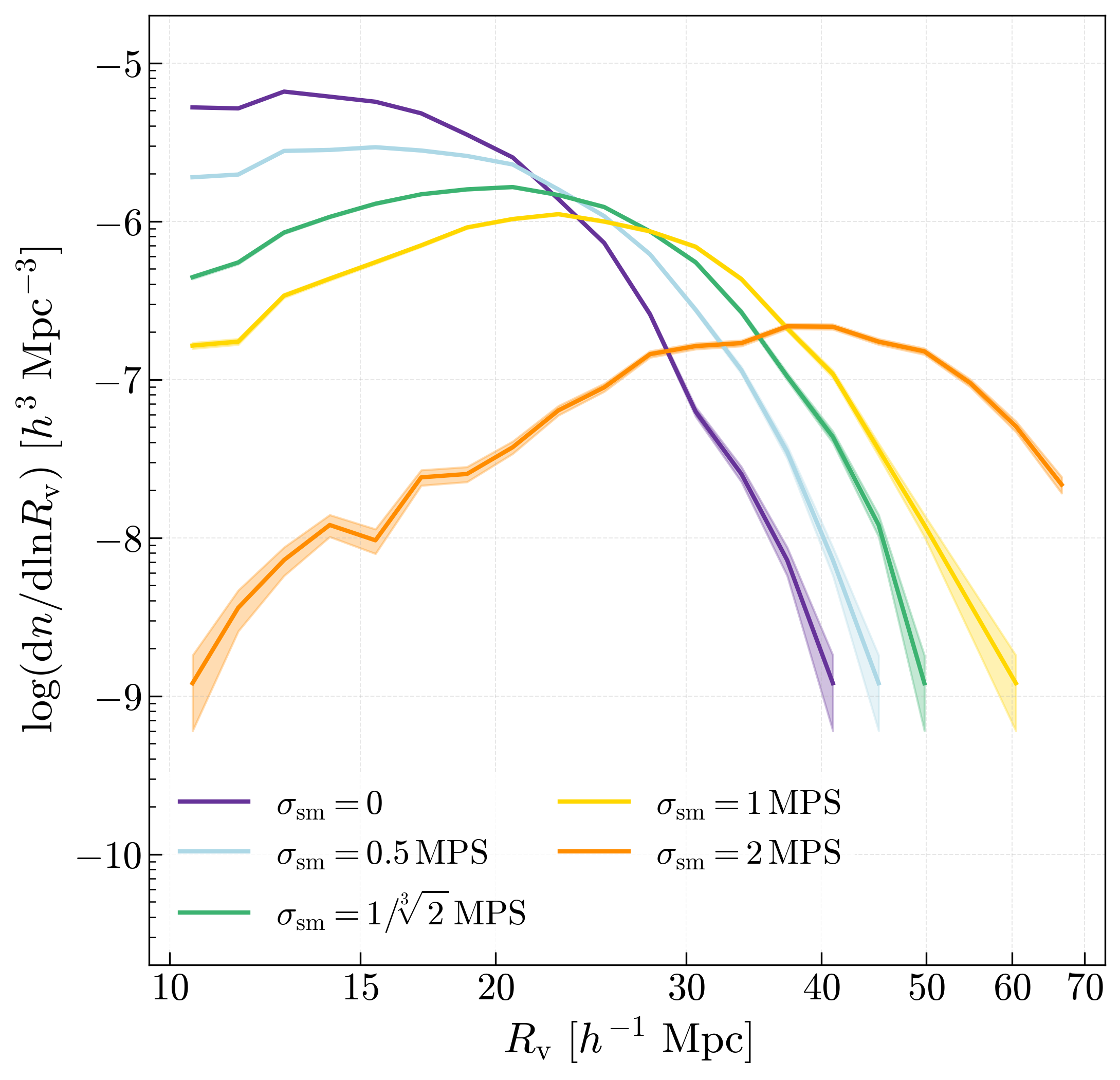}
    \caption{VSFs for different Gaussian smoothing lengths $\sigma_\mathrm{sm}$, with fixed grid resolution $1/\sqrt[3]{2}\,\mathrm{MPS}$. The associated Poissonian errors are shown as colored bands. Larger $\sigma_\mathrm{sm}$ suppresses small-scale fluctuations, enhancing larger voids; smaller $\sigma_\mathrm{sm}$ preserves fine structures, fragmenting larger voids.}
    \label{fig:vsfSM}
\end{figure}

As shown in Fig.~\ref{fig:vsfSM}, increasing $\sigma_\mathrm{sm}$ acts as a low-pass filter as expected \citep{Bond1991,SvdW2004}, suppressing small-scale fluctuations and merging nearby voids into larger ones. Smaller $\sigma_\mathrm{sm}$ preserves fine structures, often fragmenting large voids into subvoids. No single optimal value exists; $\sigma_\mathrm{sm}$ should be selected according to the target physical scale and in agreement with the purpose of the analysis.

\begin{figure} [h]
    \centering
    \includegraphics[width=0.98\linewidth]{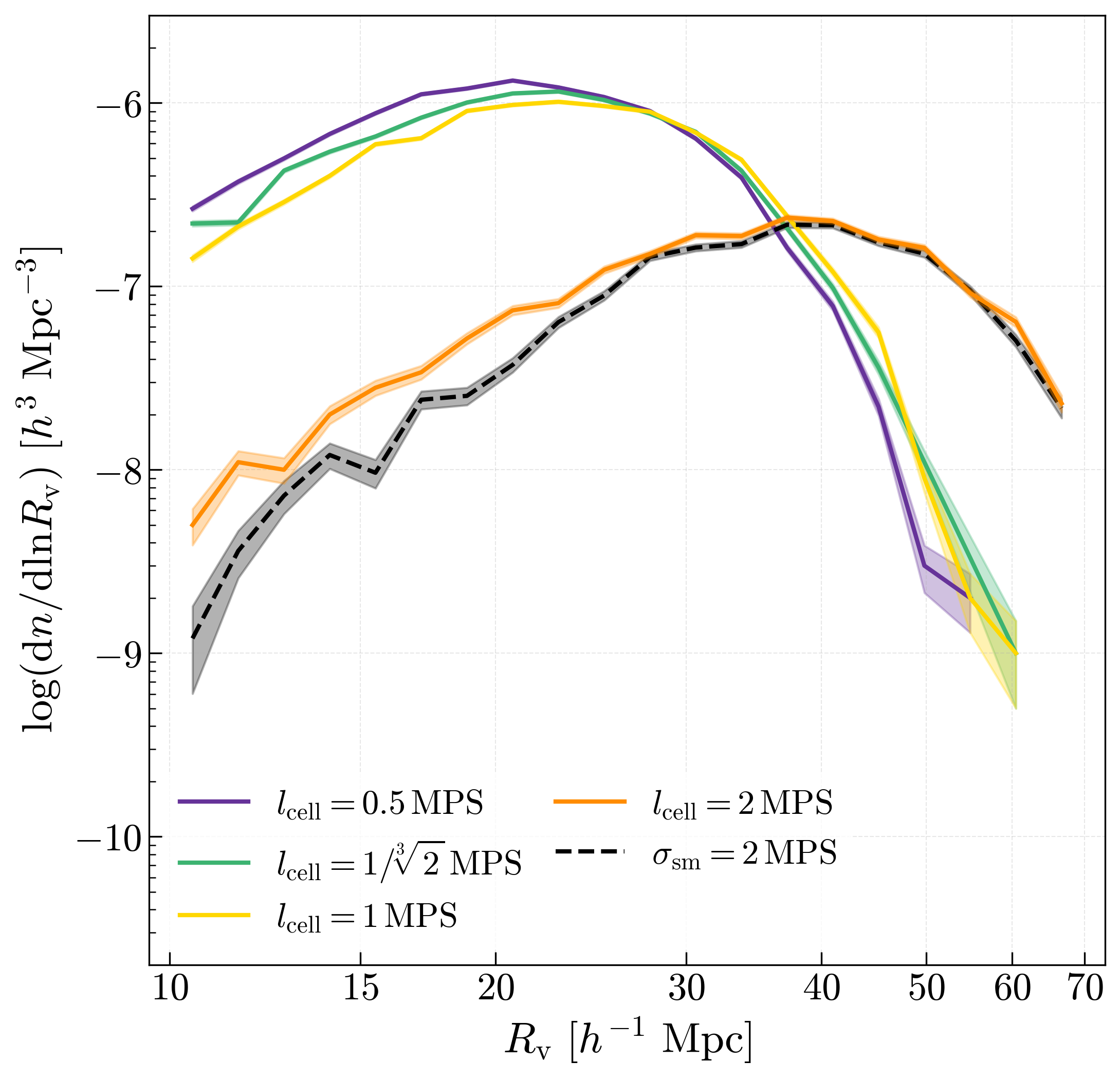}
    \caption{VSFs for different grid-cell sizes $l_\mathrm{cell}$, with fixed $\sigma_\mathrm{sm}=1\,\mathrm{MPS}$. The associated Poissonian errors are shown as colored bands. Larger cells suppress small-scale fluctuations, enhancing the detection of large voids, while smaller cells converge to a quasi-stable VSF. The VSF for $\sigma_\mathrm{sm}=2\,\mathrm{MPS}$ from Fig. \ref{fig:vsfSM}, in black, is reported for reference.}
    \label{fig:vsfCS}
\end{figure}

Finally, fixing $\sigma_\mathrm{sm}=1\,\mathrm{MPS}$, we vary the grid-cell size between $l_\mathrm{cell} = 0.5\,\mathrm{MPS}$ and $l_\mathrm{cell} =2\,\mathrm{MPS}$ (Fig.~\ref{fig:vsfCS}). Larger cells smooth the field favoring the detection of large voids, while smaller cells yield stable results once their size drops below the smoothing scale.
Notably, the VSF obtained with large cells is comparable to that derived from the void sample identified with $\sigma_\mathrm{sm} = 2\,\mathrm{MPS}$ in Fig.~\ref{fig:vsfSM}, indicating that the overall smoothing is dominated by the larger of the two scale parameters.

Reducing the cell size below the smoothing scale has negligible impact on the VSF but increases computational cost. Therefore, optimal efficiency is achieved when the grid resolution and smoothing length are comparable, ensuring both stability and minimal noise amplification.

\section{Derivations for the reconstructed-space formalism for optimal transport displacement}
\label{app:otderiv}

In this appendix, we present the main derivations underlying the \citet{Kaiser1987} redshift-to-real-space mapping, as adapted in this work to account for the linear 
tracer bias $b$ in the OT-reconstructed displacement field. This mapping is employed in the context of \texttt{BitVF}, introduced in Sect.~\ref{sec:bitvf}. In particular, we applied this formulation in Sect.~\ref{sec:recspace} to correct for the peculiar 
velocities of the Buzzard mock LRGs and recover their reconstructed real-space positions, in order to assess the effectiveness of our method in mitigating 
redshift-space distortions.

The Kaiser relation \citep{Kaiser1987} connects the real-space positions $\mathbf{r}$ and the redshift-space positions $\mathbf{s}$ of tracers through their LOS peculiar velocities:
\begin{equation}
\mathbf{r} = \mathbf{s} - \frac{v_\parallel}{aH}\hat{l} \, ,
\end{equation}
where $v_\parallel$ is the component of the peculiar velocity along the LOS $\hat{l}$, $a$ is the scale factor, 
and $H$ is the Hubble parameter. In the linear regime, the peculiar velocity field \(\mathbf{v}\) is related to the displacement field $\mathbf{\Psi}$ through
\begin{equation}
    \label{eq:velocity_app}
\mathbf{v} = a f H \mathbf{\Psi} \, ,
\end{equation}
with $f$ denoting the linear growth rate. Substituting this relation into the previous equation, the mapping becomes
\begin{equation}
\mathbf{r} = \mathbf{s} - f\mathbf{\Psi}_\parallel \hat{l} \, ,
\end{equation}
which shows explicitly how the redshift-space distortions along the LOS are proportional to the parallel component of the displacement field.

For tracers that move from their Lagrangian positions, the trajectory can be expressed as
\begin{equation}
    \label{eq:velocityExt_app}
\mathbf{x}(\mathbf{q},D) = \mathbf{q} + \mathbf{\Psi}(\mathbf{q},D) \, , \qquad
\mathbf{v} = a f D H \frac{\mathrm{d}\mathbf{\Psi}}{\mathrm{d}D} \, ,
\end{equation}
where $\mathbf{q}$ denotes the Lagrangian coordinate, $D$ is the linear growth factor, and $\mathbf{\Psi}(\mathbf{q},D)$ is the time-dependent displacement. At sufficiently large scales, in the linear regime, the velocity of the matter field is assumed to be independent of the tracer considered, i.e., unbiased:
\begin{equation}
    \label{eq:equalvel}
    \mathbf{v}_\mathrm{tr} = \mathbf{v}_\mathrm{m} \quad \Rightarrow \quad \frac{\mathrm{d}\mathbf{\Psi}_\mathrm{tr}}{\mathrm{d}D} = \frac{\mathrm{d}\mathbf{\Psi}_\mathrm{m}}{\mathrm{d}D} \, ,
\end{equation}
where the subscript $\mathrm{tr}$ refers to a generic biased tracer and $\mathrm{m}$ to the underlying matter field.

In first-order Lagrangian perturbation theory, the displacement field $\mathbf{\Psi}(\mathbf{q},D)$ is related to the density contrast in Lagrangian space $\delta(\mathbf{q},D)$ through the Poisson equation:
\begin{equation}
    \label{eq:densPsi}
    \delta(\mathbf{q},D) = - \nabla_{\mathbf{q}} \cdot \mathbf{\Psi}(\mathbf{q},D) \, .
\end{equation}
Since in linear theory the density contrast grows as \citep{Zeldovich1970}
\begin{equation}
    \delta(\mathbf{q},D) = \frac{D}{D_0} \, \delta(\mathbf{q},D_0) \, ,
\end{equation}
we can write
\begin{equation}
    \nabla_{\mathbf{q}} \cdot \mathbf{\Psi}(\mathbf{q},D) = - \frac{D}{D_0} \, \delta(\mathbf{q},D_0) \, .
\end{equation}
Taking the derivative with respect to $D$, one obtains
\begin{equation}
    \frac{\mathrm{d}}{\mathrm{d}D} \nabla_{\mathbf{q}} \cdot \mathbf{\Psi}(\mathbf{q},D) = - \frac{\delta(\mathbf{q},D_0)}{D_0} \, .
\end{equation}
This implies that the $D$-derivative of the divergence is a time-independent quantity, set uniquely by the initial matter density field. 
Therefore, the linear growth preserves universality: the displacement divergence evolves identically for all tracers once the initial conditions 
are specified, and thus the displacement field. From now on, $\mathbf{\Psi}_\mathrm{m} = \mathbf{\Psi}_\mathrm{tr} = \mathbf{\Psi}$ and 
$\mathbf{v}_\mathrm{m} = \mathbf{v}_\mathrm{tr} = \mathbf{v}$. 

In real space, the density contrast of tracers can be related to the divergence of the Lagrangian displacement field as
\begin{equation}
\nabla_{\mathbf{q}} \cdot \mathbf{\Psi} = -\frac{\delta_\mathrm{tr}^{(r)}}{b} = - \delta_\mathrm{m}^{(r)} \, ,
\end{equation}
where \(b\) is the linear tracer bias, \(\delta_\mathrm{tr}^{(r)}\) is the tracer density contrast in real space, and \(\delta_\mathrm{m}^{(r)}\) is the underlying 
matter density contrast.

Observationally, positions are measured in redshift space rather than real space. The effect of peculiar velocities on the density contrast 
can be described in Fourier space following \citet{Kaiser1987}:
\begin{equation}
\label{eq:deltas}
\hat{\delta}^{(s)}_\mathrm{tr}(\mathbf{k}) = \left(b + f \mu^2\right) \hat{\delta}^{(r)}_\mathrm{m}(\mathbf{k}) \, , \quad \text{with} \quad \mu = \hat{\mathbf{k}} \cdot \hat{l} \, ,
\end{equation}
where \(\hat{\delta}\) indicates the Fourier transform of \(\delta\), \(\hat{\mathbf{k}}\) is the unit wavevector and
\(\hat{l}\) is the LOS direction.

Using Eq.~\eqref{eq:densPsi}, it follows that in Fourier space the displacement field reads
\begin{equation}
\label{eq:psik}
\hat{\mathbf{\Psi}}(\mathbf{k}) = \frac{i \mathbf{k}}{|\mathbf{k}|^2} \hat{\delta}^{(r)}_\mathrm{m}(\mathbf{k}) \, .
\end{equation}
This relation explicitly connects the Lagrangian displacement field to the underlying matter density fluctuations, 
forming the basis for reconstructing the real-space positions of tracers from redshift-space observations.

Combining Eqs.~\eqref{eq:deltas} and \eqref{eq:psik}, we can formally write:
\begin{equation}
\hat{\mathbf{\Psi}}(\mathbf{k}) = \frac{i \mathbf{k}}{|\mathbf{k}|^2} \frac{\hat{\delta}^{(s)}_\mathrm{tr}(\mathbf{k})}{(b+f\mu^2)} \quad \Longrightarrow \quad b\left(1+\frac{f}{b}\mu^2\right) \hat{\mathbf{\Psi}}(\mathbf{k}) = \frac{i \mathbf{k}}{|\mathbf{k}|^2} \hat{\delta}^{(s)}_\mathrm{tr}(\mathbf{k}) \, .
\end{equation}
Taking the divergence of both sides and transforming back to configuration space yields:
\begin{equation}
\label{eq:nablaInter_app}
\nabla_{\mathbf{q}} \cdot \mathbf{\Psi} + \frac{f}{b} \nabla_{\mathbf{q}} \cdot \big[ \hat{l} (\hat{l} \cdot \mathbf{\Psi}) \big] = - \frac{\delta^{(s)}_\mathrm{tr}}{b} \, .
\end{equation}
The OT reconstruction provides, by definition,
\begin{equation}
\label{eq:nablaOT_app}
\nabla_{\mathbf{q}} \cdot \mathbf{\Psi}_\mathrm{OT} = - \delta^{(s)}_\mathrm{tr} \, ,
\end{equation}
where the subscript OT indicates the displacement obtained from our reconstruction.
Eq.~\eqref{eq:nablaInter_app} can be rewritten as:
\begin{equation}
b \nabla_{\mathbf{q}} \cdot \Big( \mathbf{\Psi} + \frac{f}{b} \mathbf{\Psi}_\parallel \Big) = \delta^{(s)}_\mathrm{tr} \, ,
\end{equation}
which, combined with Eq.~\eqref{eq:nablaOT_app}, gives
\begin{equation}
\label{eq:otpsi}
\mathbf{\Psi}_\mathrm{OT} = b\ \Big( \mathbf{\Psi} + \frac{f}{b} \mathbf{\Psi}_\parallel \Big) \, .
\end{equation}
Projecting Eq.~\eqref{eq:otpsi} along the LOS, we obtain:
\begin{equation}
\label{eq:psipar_app}
\mathbf{\Psi}_{\mathrm{OT}\parallel} = (b+f) \mathbf{\Psi}_\parallel \quad \Longrightarrow \quad \mathbf{\Psi}_\parallel = \frac{\mathbf{\Psi}_{\mathrm{OT}\parallel}}{(b+f)} \, .
\end{equation}
Knowing that the displacement scales with the linear growth factor \(D\) as
\begin{equation}
\mathbf{\Psi}(D) = \frac{D}{D_0} \mathbf{\Psi}(D_0) \, ,
\end{equation}
we then have
\begin{equation}
\frac{\mathrm{d}\mathbf{\Psi}(D)}{\mathrm{d}D} = \frac{1}{D_0} \mathbf{\Psi}(D_0) \, .
\end{equation}
Using this relation in Eq.~\eqref{eq:velocityExt_app}, we recover the formulation for \(\mathbf{v}\) as expressed in Eq.~\eqref{eq:velocity_app}.

Considering only the LOS component of \(\mathbf{v}\) and using Eq.~\eqref{eq:psipar_app}, we find
\begin{equation}
\mathbf{v}_\parallel = a H \frac{f}{(b+f)} \mathbf{\Psi}_{\mathrm{OT}\parallel} \, .
\end{equation}
This result provides the fundamental relation connecting the LOS velocity, responsible for RSD, with the OT displacement. 
The factor \(1/(b+f)\) accounts for the linear bias introduced by using the OT reconstruction. Finally, the mapping from redshift to real space 
using the biased OT displacement is given by
\begin{equation}
\label{OTrspaceCorr_app}
\mathbf{r} = \mathbf{s} - \frac{f}{(b+f)} \mathbf{\Psi}_{\mathrm{OT}\parallel} \hat{l} \, .
\end{equation}

We note that the displacement recovered by our OT reconstruction algorithm is backward in time. Therefore, in practical applications, 
the sign of the reconstructed displacement field must be reversed to properly correct for RSD.

\label{LastPage}
\end{document}